\documentclass{article}

\usepackage{times}

\usepackage{amsmath}
\usepackage{amssymb}
\usepackage{amsfonts}
\usepackage{mathrsfs}

\usepackage{graphicx}
\usepackage{geometry}

\usepackage[breaklinks=true]{hyperref}
\usepackage[backend=bibtex,style=numeric,sorting=none]{biblatex}

\usepackage{appendix}

\bibliography{references}

\begin{document}
    \hypersetup{pageanchor=false}
    \begin{titlepage}
      
        \begin{flushright}
          {\bf IFJPAN-IV-2024-12} 
        \end{flushright}

        \vspace*{30mm}
        \begin{center}
            \huge{
                \textbf{
                    Pseudo-observables and  Deep Neural Network for mixed CP
                    $H \to \tau \tau$ decays at LHC.
                }
            }
        
            \vspace{1.5cm}
            \fontsize{10pt}{12pt}\selectfont
            \textbf{E. Richter-Was$^{a}$, T. Yerniyazov$^{b}$ and Z. Was$^{c}$}

            \vspace{0.5cm}
            \fontsize{10pt}{12pt}\selectfont
            \textit{$^a$ Institute of Physics, Jagiellonian University, Łojasiewicza 11, 30-348 
                Kraków, Poland}
            
            \textit{$^b$ Faculty of Physics, Astronomy and Applied Computer Science, Jagiellonian 
                University, Łojasiewicza 11, 30-348 Kraków, Poland}
            
            \textit{$^c$ Institute of Nuclear Physics, IFJ-PAN, Radzikowskiego 152, 31-342 Kraków, 
                Poland}
        \end{center}

        \vfill
        \begin{center}
            \textbf{ABSTRACT}
        \end{center}
        The consecutive steps of cascade decay initiated by $H\to \tau\tau$ can be useful for the 
        measurement of Higgs couplings and in particular of the Higgs boson parity. The standard 
        analysis method applied so far by ATLAS and CMS Collaborations was to fit a one-dimensional 
        distribution of the angle between reconstructed decay planes of the $\tau$ lepton decays, 
        $\phi^*$, which is sensitive to transverse spin correlations of the $\tau$ decays and,
        hence, to the CP state mixing angle, $\phi^{CP}$.

        Machine Learning techniques (ML) offer opportunities to manage such complex 
        multidimensional signatures. The  multidimensional signatures, the 4-momenta of the $\tau$ 
        decay products can be used directly as input information to the machine learning algorithms 
        to predict the CP-sensitive pseudo-observables and/or provide discrimination between 
        different CP hypotheses. In the previous papers, we have shown that multidimensional 
        signatures, the 4-momenta of the $\tau$ decay products can be used  as input 
        information to the machine learning algorithms to predict the CP-sensitive 
        pseudo-observables and/or provide discrimination between different CP hypotheses.

        In this paper we show how the classification or regression methods can be used to train an 
        ML model to predict the spin weight sensitive to the CP state of the decaying Higgs boson, 
        parameters of the functional form of the spin weight, or the most preferred CP mixing angle 
        of the analysed sample. The one-dimensional distribution of the predicted spin weight or 
        the most preferred CP mixing angle of the experimental data can be examined further, with
        the statistical methods, to derive the measurement of the CP mixing state of the Higgs signal
        events in decay $H\to \tau\tau$.

        This paper extends studies presented in our previous publication, with more experimentally
        realistic scenarios for the features lists and further development of strategies how 
        proposed pseudo-observables can be used in the measurement. 

        \vfill
        
        \small
        \begin{flushleft}
          {\bf IFJPAN-IV-2024-12} \\
            November 2024
        \end{flushleft}
        
        \footnoterule
        \footnotesize
        T. Y. was supported in part by grant No. 2019/34/E/ST2/00457 of the National Science Center 
        (NCN), Poland and also by the Priority Research Area Digiworld under the program Excellence 
        Initiative Research University at the Jagiellonian University in Cracow.

        The majority of the numerical calculations were performed at the PLGrid Infrastructure of 
        the Academic Computer Centre CYFRONET AGH in Kraków, Poland.
    \end{titlepage}
    \hypersetup{pageanchor=true}

    \section{Introduction}

        Machine learning techniques are more and more often being applied in high-energy physics 
        data analysis. With Tevatron and LHC experiments, they became an analysis standard. 
        For the recent reviews see e.g.~\cite{Guest:2018yhq,Carleo:2019ptp,Albertsson:2018maf}.

        In this paper, we present how ML techniques can help exploit the substructure of the 
        hadronically decaying $\tau$ leptons in the measurement of the Higgs boson CP-state mixing 
        angle $\phi^{CP}$ in $H \to \tau \tau$ decay. This problem has a long 
        history~\cite{Kramer:1993jn,Bower:2002zx}, and was studied both for 
        electron-positron~\cite{Rouge:2005iy,Desch:2003mw} and for 
        hadron-hadron~\cite{Berge:2008dr,Berge:2015nua} colliders. Following the proposed by those 
        papers one-dimensional observable, a polar angle $\phi^{*}$ between $\tau$ leptons decay 
        planes, ATLAS and CMS Collaborations measured the CP mixing angle $\phi^{CP}$ with Run 2 
        data collected in pp collisions at LHC. The  Standard Model (SM) predicts that the scalar 
        Higgs boson is in the  $\phi^{CP} = 0^{\circ}$ state while the pseudo-scalar would be in the 
        $\phi^{CP} = 90^{\circ}$ state for the Higgs couplings to $\tau$ pairs. The measured value
        by ATLAS Collaboration ~\cite{ATLAS:2022akr} is $\phi^{CP} = 9^{\circ}\pm 16^{\circ}$, and 
        by CMS Collaboration ~\cite{CMS:2021sdq}  $\phi^{CP} = -1^{\circ}\pm 19^{\circ}$. While data
        disfavour pure pseudo-scalar coupling with more than 3 standard deviations significance and 
        are compatible with SM expectations of Higgs couplings to $\tau$ leptons being scalar-like 
        ($\phi^{CP}$ = 0), there is still room for improvement of the precision of the measurement
        with more data collected during ongoing LHC Run 3 and then even 10 times more data 
        expected with High-Lumi LHC. 

        The theoretical basis for the  measurement is simple, the cross-section dependence on the 
        parity mixing angle has the form of the second-order trigonometric polynomial in the 
        $\phi^{CP}$ angle. It can be implemented in the Monte Carlo simulations as the per event 
        spin weight $wt$, which parametrises this sensitivity, see~\cite{Przedzinski:2014pla} for 
        more details. Even if the problem in its nature is multidimensional: correlations between
        direction and momenta of the $\tau$ lepton decay products, the ML has not been used so far 
        for experimental analysis design. This is in part, because ML adds complexity to the data 
        analysis and requires relatively high statistics of the signal events. Also, the ML 
        solutions need to be evaluated for their suitability in providing estimates of systematic 
        ambiguities.

        In~\cite{Jozefowicz:2016kvz, Lasocha:2018jcb} we have presented ML-based analysis on the 
        Monte Carlo simulated events for the three channels of the $\tau$ lepton-pair decays: 
        $\rho^\pm\nu_\tau\rho^\mp\nu_\tau$, $a_1^\pm\nu_\tau\rho^\mp\nu_\tau$ and 
        $a_1^\pm\nu_\tau a_1^\mp\nu_\tau$. In those papers we limited the scope to the 
        scalar-pseudoscalar classification case. We  explored the kinematics of outgoing decay 
        products of the $\tau$ leptons and geometry of decay vertices which can be used for 
        reconstructing original $\tau$ lepton directions. In the following 
        paper~\cite{Lasocha:2020ctd}, we extended ML analysis to a measurement of the 
        scalar-pseudoscalar mixing angle $\phi^{CP}$ of the $H\to\tau\tau$ coupling, limiting it, 
        however, to the ideal situation  where the 4-momenta of all outgoing $\tau$ decay products 
        were known. In the experimental conditions, this is not the case, because of the two 
        neutrinos from $\tau$ decays which cannot be measured directly.  

        In this paper, we extend further our research to more experimentally realistic 
        representations of $H \to \tau \tau$ events and focus on using the per-event spin weight 
        $wt$ as a one-dimensional pseudo-observable. We train a machine learning model to predict 
        those weights and relay on for further statistical analysis to infer the $\phi^{CP}$ mixing 
        angle from predicted distribution. This pseudo-observable can be used as 
        an alternative or complementary to the angle $\phi^*$ between $\tau$ lepton decay planes. 
        As in~\cite{Lasocha:2020ctd} we still constrain ourselves to the measurement of the coupling
        in the most promising decay channel $H \to \tau^+ \tau^- \to \rho^+ \nu_\tau \rho^-\nu_\tau
        \to \pi^+ \pi^0 \nu_\tau \pi^- \pi^0 \nu_\tau $.

        We analyse potential solutions using {\it deep learning } 
        algorithms~\cite{dlbook} and a {\it Deep Neural Network} ({\it DNN}) implemented in the 
        {\it Tensorflow} environment~\cite{tensorflow2015-whitepaper}, previously found effective 
        for binary classification~\cite{Jozefowicz:2016kvz, Lasocha:2018jcb} between scalar and 
        pseudoscalar Higgs boson hypotheses.

        We employ the same DNN architecture used for multiclass classification and regression 
        in~\cite{Lasocha:2020ctd}. The DNN is tasked with predicting per-event:
        
        \begin{itemize}
            \item
                Spin weight as a function of the CP mixing angle $\phi^{CP}$.
            \item
                Decay configuration dependent coefficients, for the functional form of the 
                spin weight distribution as a function of the mixing angle $\phi^{CP}$.
            \item
                The most preferred mixing angle, i.e. the value from which the spin weight reaches 
                its maximum.
        \end{itemize}

        These goals are complementary or even to a great extent redundant, e.g. with the functional 
        form of the spin weight. We can easily find the mixing angle at which it has a maximum. 
        However, the precision of predicting that value may not be the same for 
        different methods. All three cases are examined as classification and regression problems.
        
        The paper is organised as follows: Section \ref{Sec:CPtheory} describes the basic 
        phenomenology of the problem. Properties of the matrix elements are discussed and the 
        notation used is introduced. In Section~\ref{Sec:Features} we review feature sets 
        (lists of variables) used as an input to the DNN and present samples prepared for analyses. 
        Section~\ref{Sec:Training-Variant-All} outlines ML methods and the performance of the
        DNN. Section~\ref{Sec:Training-Variant-XXX} delves deeper into the performance across the
        feature sets with varying levels of realism. Section~\ref{Sec:Results} presents the final 
        discussion on pseudo-observables resulting from the trained DNN and prospects for using 
        them for experimental measurement. Section~\ref{Sec:Summary} (Summary) closes the paper.

        In Appendix~\ref{App:DNN} more technical details on the DNN architecture are given, 
        and input data formulation for classification and regression cases is explained.
        We also collect figures monitoring the performance of DNN training.

    \section{Physics description of the problem}
        \label{Sec:CPtheory}
        The most general Higgs boson Yukawa coupling to a $\tau$ lepton pair, expressed with the 
        help of the scalar--pseudoscalar parity mixing angle $\phi^{CP}$ reads as
        
        \begin{equation}
            {\cal L}_Y= N\;\bar{\tau}{\mathrm  h}(\cos\phi^{CP}+i\sin\phi^{CP}\gamma_{5})\tau,
        \end{equation}
        
        where $N$ denotes normalisation, $\mathrm h$ Higgs field and $\bar\tau$, $\tau$ spinors of 
        the $\tau^+$ and $\tau^-$. As we will see later, this simple analytic form translates itself
        into useful properties of observable distributions convenient for our goal of determining 
        $\phi^{CP}$. Recalling these definitions is therefore justified, aiding in the 
        systematisation of observable quantity (feature) properties and correlations.

        The squared matrix element for the scalar/pseudoscalar/mixed parity Higgs, decaying into
        $\tau^+ \tau^-$ pairs can be expressed as
        
        \begin{equation} \label{eq:matrix}
            |M|^2\sim 1 +  h_{+}^{i}  h_{-}^{j} R_{i,j}; \;\;\;\;\; i,j=\{x,y,z\}
        \end{equation}
        
        where  $h_{\pm}$ denote polarimetric vectors of $\tau$ decays (solely defined by $\tau$ 
        decay matrix elements) and $R_{i,j}$ is the density matrix of the $\tau$ lepton pair spin 
        state. Details of the frames used for $R_{i,j}$ and $h_{\pm}$ definition are given in 
        Ref.~\cite{Desch:2003rw}. The corresponding CP-sensitive spin weight $wt$ has the form:
        
        \begin{equation}
            \label{eq:wt_master}
            wt = 1-h_{{+}}^{z} h_{{-}}^{z}+ h_{{+}}^{\perp} R(2\phi^{CP})~h_{{-}}^{\perp}.
        \end{equation}

        The formula is valid for $h_\pm$ defined in $\tau^\pm$ rest-frames, $h^{z}$ stands for 
        longitudinal and  $h^{\perp}$ for transverse components of $h$. The $R(2\phi^{CP})$ denotes 
        the $2\phi^{CP}$ angle rotation matrix around the $z$ direction:
            $R_{xx}= R_{yy}={\cos2\phi^{CP}}$, $R_{xy}=-R_{yx}={\sin2\phi^{CP}}$.
        The $\tau^\pm$ decay polarimetric vectors $h_{+}^i$,  $h_{-}^j$, in the case of 
        $\tau^{\pm} \to \pi^{\pm} \pi^0 \nu $ decay, read 
        
        \begin{equation}
        \label{eq:h_master}
            h^i_\pm =  {\cal N} \Bigl( 2(q\cdot p_{\nu})  q^i -q^2  p_{\nu}^i \Bigr), \;\;\;  
        \end{equation}
        
        where $\tau$ decay products $\pi^{\pm}$, $\pi^0$ and $\nu_{\tau}$ 4-momenta are denoted 
        respectively as $ p_{\pi^{\pm}}$, $p_{\pi^0}, p_{\nu}$ and $q=p_{\pi^\pm} -p_{\pi^0}$.
        Obviously, complete CP sensitivity can be extracted only if $p_{\nu}$ is known.

\begin{figure}
            \begin{center} {
                \includegraphics[width=5.0cm,angle=0]{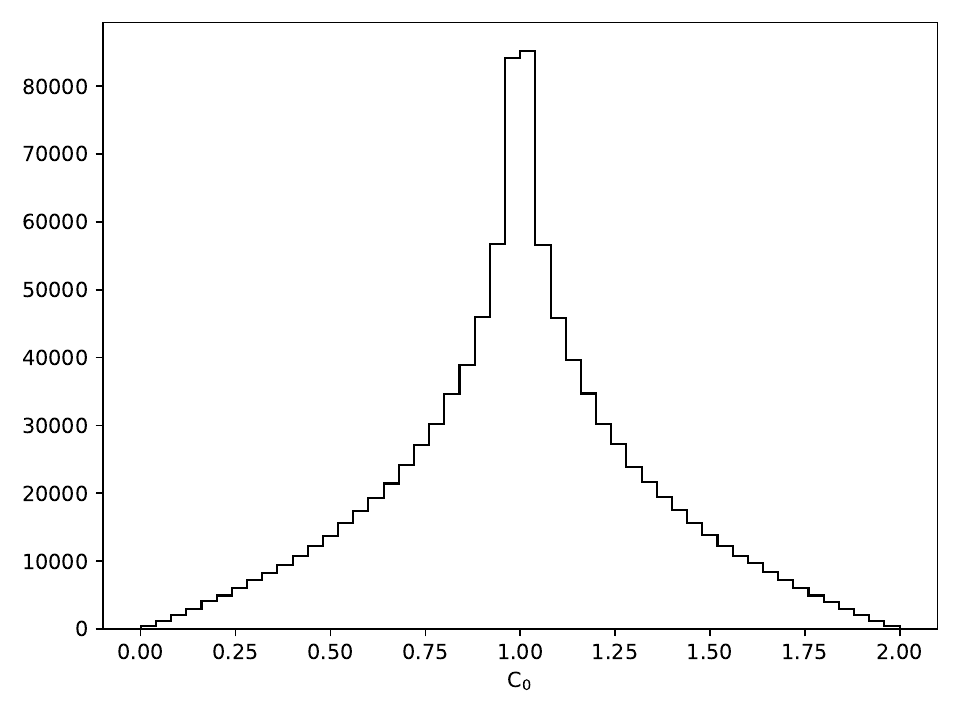}
                \includegraphics[width=5.0cm,angle=0]{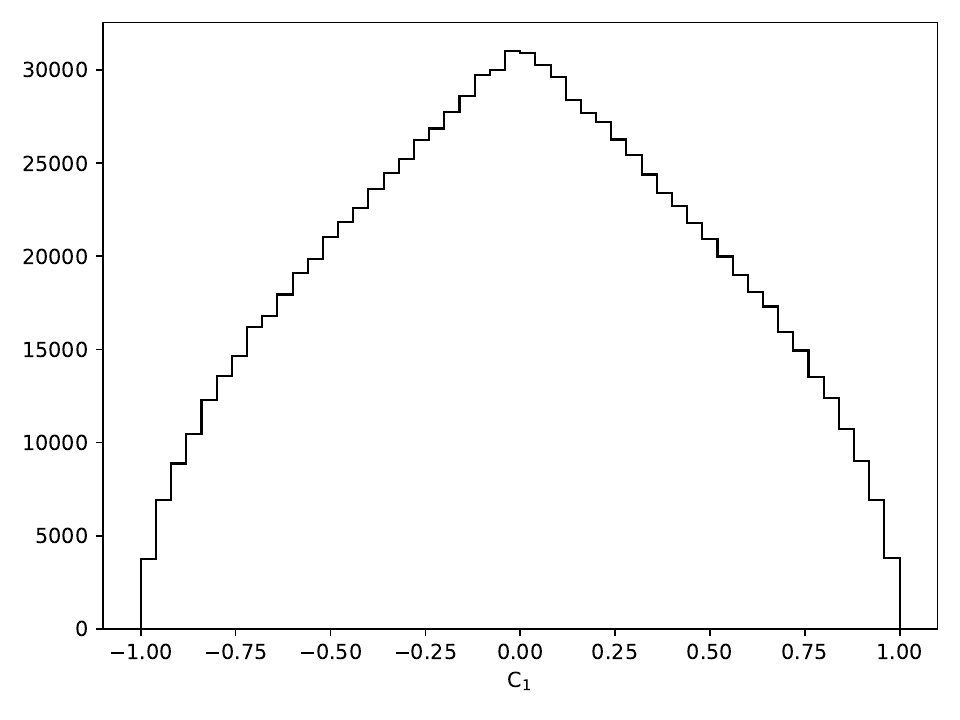}
                \includegraphics[width=5.0cm,angle=0]{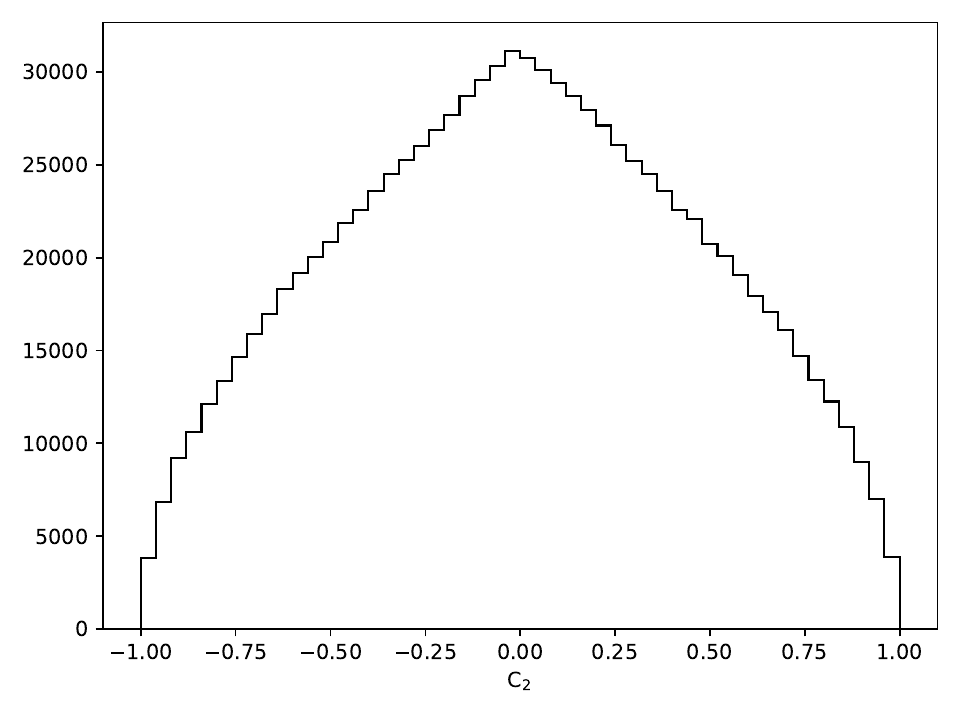}
            }
            \end{center}
            \caption{Distributions of the $C_0, C_1, C_2$ coefficients of 
                formula~(\ref{eq:ABC_alpha}), for the $H \to \tau \tau$ events sample.}
            \label{fig:c012s}
        \end{figure}

        Note that the spin weight $wt$ is a simple first-order trigonometric polynomial in a 
        $2 \phi^{CP}$ angle. This observation  is valid for all $\tau$ decay channels. For the 
        clarity of the discussion on the DNN results, we introduce $\alpha^{CP}= 2\phi^{CP}$,
        which spans over $(0, 2\pi)$ range. $\alpha^{CP}= 0, 2\pi$ corresponds to the scalar 
        state, while $\alpha^{CP}= \pi$ corresponds to the pseudoscalar one. The spin weight can be 
        expressed as

        \begin{equation}
            wt = C_0 + C_1 \cdot \cos(\alpha^{CP}) + C_2 \cdot \sin(\alpha^{CP}),
            \label{eq:ABC_alpha}
        \end{equation}

        where coefficients
        
        \begin{eqnarray}
            \label{eq:CPcoeff}
            C_0 &=& 1 -  h_{+}^{z} h_{-}^{z}, \nonumber \\
            C_1 &=& - h_{+}^{x} h_{-}^{x} +  h_{+}^{y} h_{-}^{y}, \\
            C_2 &=& - h_{+}^{x} h_{-}^{y} -  h_{+}^{y} h_{-}^{x},\nonumber 
        \end{eqnarray}
    
        depend on the $\tau$ decays only. Distribution of the $C_0, C_1, C_2$ coefficients, 
        for the sample of $H \to \tau \tau$ events used for our numerical results is shown in 
        Figure~\ref{fig:c012s}. The $C_0$ spans $(0, 2)$ range, while $C_1$ and $C_2$ of (-1, 1) 
        range have a similar shape, quite different than the one of $C_0$.

        The amplitude of the spin weight $wt$ as a function of $\alpha^{CP}$ depends on the 
        multiplication of the length of the transverse components of the polarimetric vectors. The 
        longitudinal component $h_{+}^{z} h_{-}^{z}$ defines shift with respect to zero of the $wt$ 
        mean value and is constant over a full $(0, 2\pi)$ range of $\alpha^{CP}$. The $wt$ 
        distribution maximum is reached for $\alpha^{CP}$ = $\measuredangle (h_{+}^{T}, h_{-}^{T})$,
        the opening angle of the transverse components of the polarimetric vectors.

        The spin weight $wt$ of the formula (\ref{eq:ABC_alpha}) can be used to introduce transverse 
        spin effects  into the event sample when for the generation transverse spin effects were not
        taken into account at all. The above statement holds true regardless of whether longitudinal 
        spin effects were included and which $\tau$ decay channels complete cascade of 
        $H \to \tau \tau$ decay. The shape of weight dependence on the Higgs coupling to the $\tau$ 
        parity mixing angle is preserved.

\begin{figure}
            \begin{center} {
                \includegraphics[width=16.0cm,angle=0]{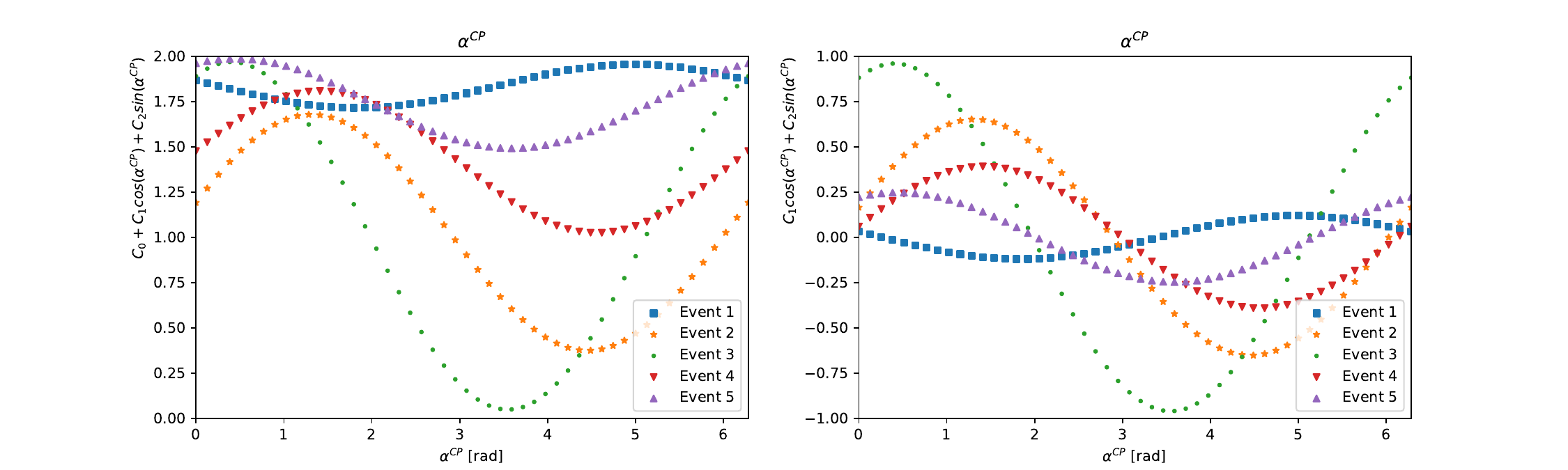}
            }
            \end{center}
            \caption{The spin weight $wt$ (left plot) and only its $\alpha^{CP}$ dependent component
                (right plot) for a few $H \to \tau \tau$ events. Note the vertical scale difference 
                between the plots due the inclusion of the $C_0$ term in the left plot and its 
                absence in the right.
            \label{fig:CPweight}}
        \end{figure}

        In Fig.~\ref{fig:CPweight} we show the distribution of the spin weight $wt$ for five example 
        $H \to \tau \tau$ events. For each event, a single value of $\alpha^{CP}$ is the most 
        preferred (corresponding to the largest weight), determined by its polarimetric vectors.

    \section{Monte Carlo samples and feature sets}
        \label{Sec:Features}

        For compatibility with our previous publications~\cite{Jozefowicz:2016kvz, Lasocha:2018jcb, 
        Lasocha:2020ctd}, we use the same samples of generated events, namely Monte Carlo events of 
        the Standard Model, 125~GeV Higgs boson, produced in pp collision at 13 TeV centre-of-mass 
        energy, generated with {\tt Pythia 8.2}~\cite{Sjostrand:2014zea}, with $\tau$ leptons decays
        simulated with the {\tt Tauolapp} library~\cite{Davidson:2010rw}. Samples are generated 
        without spin correlations included, each $\tau$ lepton is decayed independently.

        The spin effects are introduced with the spin weight $wt$ calculated with the
        {\tt TauSpinner}~\cite{Czyczula:2012ny, Przedzinski:2014pla, Przedzinski:2018ett} package,
        for several different hypotheses of the CP mixing angle $\alpha^{CP}_i$ and stored together 
        with information on  $\tau$ leptons decay products. The spin weight (\ref{eq:wt_master}),
        is calculated using $R_{i,j}$ density matrix and  polarimetric vectors $h_\pm$, as explained
        in the previous Section. In the process of the analysis, for a given event it is possible to
        restore the information of the per-event coefficients $C_0, C_1, C_2$, using its value of 
        the spin weight $wt$ at three $\alpha^{CP}$ hypotheses and solving the linear 
        equation~(\ref{eq:ABC_alpha}) or fitting the ~(\ref{eq:ABC_alpha}) formula.

        In this paper we present results for the case when both $\tau$ decays $\tau^{\pm} \to 
        \rho^{\pm} \nu_{\tau}$ and about $10^6$ simulated Higgs events are used. To partly emulate 
        detector conditions, a minimal set of cuts is used. We require that the transverse momenta 
        of the visible decay products combined, for each $\tau$, are larger than 20 GeV. It is also 
        required that the transverse momentum of each $\pi^{\pm}$ is larger than 1 GeV.

        The emphasis of the paper is to explore the feasibility to learn spin weight $wt$ or its
        components $C_0, C_1, C_2$  with ML techniques. For the ML training as feature sets 
        (quantities) we consider few scenarios introduced in paper~\cite{Lasocha:2018jcb}. We discuss
        the case of an idealistic benchmark scenario with the {\tt Variant-All} feature set, where 
        the four-momenta of {\it all} decay products of $\tau$ leptons are defined in the rest frame 
        of intermediate resonance pairs ($\rho-\rho$ system). Scenarios which are more realistic in 
        experimental conditions rely on the information that could be reconstructed from 
        experimental data. The first one, {\tt Variant-1.1}, is based on the 4-vectors and their 
        products for visible decay products only. The second one, {\tt Variant-4.1}, includes some 
        approximate information on the original $\tau$ lepton direction.

        Table~\ref{Tab:ML_variants} details the feature sets of the considered models. For more 
        details on the specific frames and approximations used for representing neutrinos or $\tau$ 
        lepton 4-momenta, please refer to~\cite{Lasocha:2018jcb}. We do not introduce any additional
        energy or momentum smearing, leaving the evaluation of its impact to experimental analysis.
        
        \begin{table}
            \vspace{2mm}
            \caption{
                The feature sets used as inputs to ML algorithms, categorised by {\tt Variant-X.Y}
                family. The third column number indicates the number of features for the 
                $\rho^{\pm}-\rho^{\mp}$. All 4-vectors components are calculated in the rest frame 
                of the hadronic decay products.
                \label{Tab:ML_variants}
            }
            \begin{center}
                \begin{tabular}{|l|l|c|l|}
                    \hline
                    Notation & Features & Counts & Comments \\
                    \hline
                    Variant-All & 4-vectors ($\pi^\pm, \pi^0, \nu$) & 24 & \\
                    \hline
                    Variant-1.1 & 4-vectors ($\pi^\pm, \pi^0, \rho^\pm$), $m_i^2, m_k^2, y_i, y_k, 
                    \phi^{*}_{i,k}$  & 29 & \\
                    \hline
                    Variant-4.1 & 4-vectors ($\pi^\pm, \pi^0$), 4-vectors $\tau$ & 24 & 
                    Approx.$p_{\tau} $\\
                    \hline
                \end{tabular}
            \end{center}
        \end{table}	
    
    \section{Training and evaluation: {\it Variant-All} }
        \label{Sec:Training-Variant-All}

        We start by discussing results for an idealistic case, i.e. the {\it Variant-All} feature 
        set. This reminds us to a great extent what was previously reported in the scope of 
        \cite{Lasocha:2020ctd}.

        \subsection{Learning spin weights \texorpdfstring{$wt$}{wt}}

            The DNN classifier is trained on a per-event feature set to predict an
            $N_{class}$-dimensional per-event vector of spin weights normalised to probability.
            Each component of the $wt^{norm}(\alpha^{CP})$ vector is a weight associated with a 
            discrete value of the mixing angle $\alpha^{CP}_i$. $N_{class}$ represents the number of 
            discrete points into which the range $(0, 2 \pi)$ is divided. To guarantee that 
            $\alpha^{CP}$ values of $0$, $\pi$ and $2\pi$ (corresponding to scalar, pseudoscalar,
            and again scalar states, respectively) are always included as distinct classes, the total 
            number of classes, $N_{class}$, is maintained as an odd number. As baseline, the DNN
            is trained with $N_{class}$ set to 51. 

            Similarly, in the regression case, the DNN is trained on per-event feature sets 
            with the corresponding spin weight vector, $wt_i$, for discrete $\alpha^{CP}$ values as
            the target variable. The DNN learns both shape and normalisation of the $wt$
            vector during the training.
            
\begin{figure}
                \begin{center} {
                    \includegraphics[width=16.0cm,angle=0]{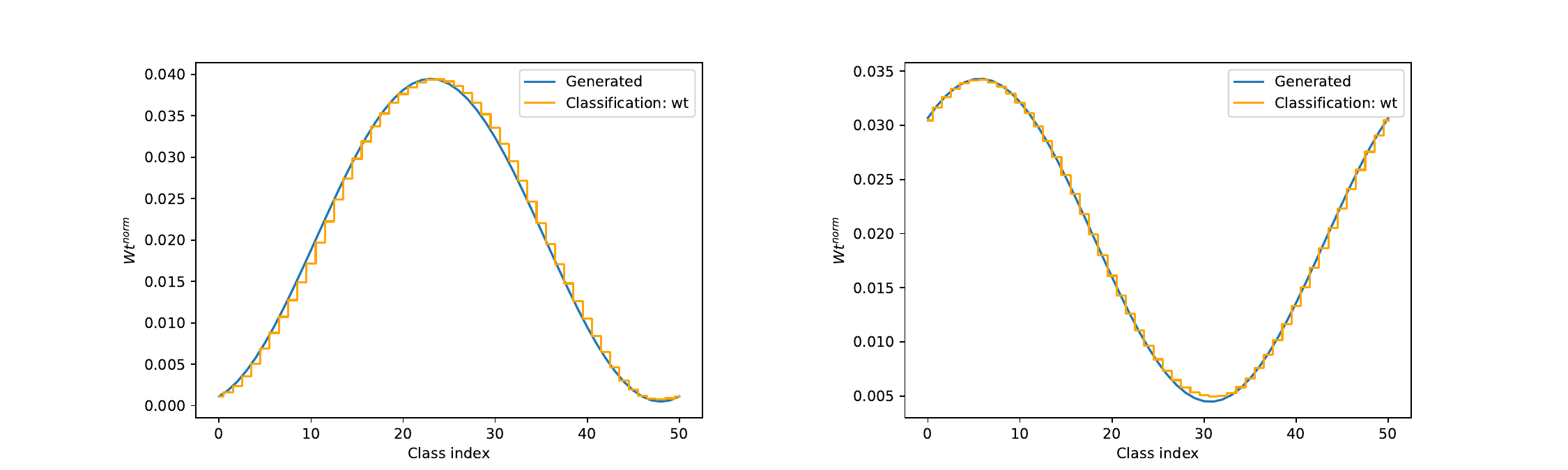}
                }
                \end{center}
                \caption{Normalised spin weight, $wt^{norm}$, predicted (orange line) and true 
                    (blue line), as a function of the class index for two example events. The
                    DNN was trained using the classification method with $N_{class}$ = 51 and 
                    {\it Variant-All}.
                }
                \label{fig:DNN_soft_nc_predwt}
            \end{figure}

            We quantify the DNN performance using physics-relevant criteria. The primary metric is 
            the ability of the DNN to reproduce the per-event spin weight, $wt^{norm}$. 
            Figure~\ref{fig:DNN_soft_nc_predwt} illustrates the true and predicted $wt^{norm}$ 
            distributions for two example events as a function of the class index $i$ (representing 
            the discretised mixing parameter $\alpha^{CP}_i$). The blue line indicates true spin 
            weights, while the orange line represents weights predicted by the DNN classifier. 
            Generally, the DNN accurately predicts the spin weight for some events, but struggles 
            for others. Encouragingly, the predicted weights exhibit the expected smooth, linear 
            combination of $\cos (\alpha^{CP})$ and $\sin (\alpha^{CP})$. This is notable given that
            the loss function does not explicitly correlate nearby classes, indicating the capacity 
            of the DNN to learn this pattern during training.
 
            The second criterion is the difference between the most probable predicted class and the
            most probable true class, denoted as $\Delta_{class}$. When calculating the difference 
            between class indices, the periodicity of the functional form 
            (Equation~(\ref{eq:ABC_alpha})) is considered. Class indices represent discrete values 
            of $\alpha^{CP}$ within the range $(0, 2\pi)$. The distance between the first and the 
            last class is zero. We compute the distance corresponding to the smaller angle 
            difference and assign a sign based on the clockwise orientation relative to the class 
            index where the true $wt$ reaches its maximum.

            From a physics perspective, learning the shape of the $wt$ distribution as a function of
            $\alpha^{CP}$ is equivalent to learning the components of the polarimetric vectors.
            Note Ref.~\cite{Cherepanov:2018yqb} for the possible extension of the methods for other
            $\tau$ decay channels. For learning the polarimetric vectors the much larger $Z \to \tau \tau$ sample
            can be used too.
           
            However, since only the shape, not the normalisation, is available, the $C_i$ 
            coefficients cannot be fully recovered from $wt^{norm}$. This is not the primary goal, 
            though. The physics interest lies in determining the preferred $\alpha^{CP}$ value for
            events in the analysed sample, i.e. the value where the summed $wt$ distribution reaches
            its maximum. This alings with the objective of potential measurement, namely, 
            determining the CP mixing of the analysed sample, which will be discussed further in 
            Section~\ref{Sec:Results}. 

            Figure~\ref{fig:DNN_delta_argmax}, top-left panel, displays the distribution
            of $\Delta_{class}$ for $N_{class}$ = 51 for the DNN trained with the 
            classification method and {\it Variant-All/4.1/1.1}. The distribution is Gaussian-like and 
            centered around zero with standard deviation $\sigma_{\Delta_{class}}$ = 0.179 [rad]
            for {\it Variant-All}.
            The top-right panel of Figure~\ref{fig:DNN_delta_argmax} shows the 
            distribution of $\Delta_{class}$ for $N_{class}$ = 51 using the regression method. 
            In this case, $\sigma_{\Delta_{class}}$ = 0.170 [rad] for {\it Variant-All}. 
            Our findings indicate that the accuracy of learning the spin weight $wt$ using 
            classification and regression methods is comparable. 

        \subsection{Learning \texorpdfstring{$C_0, C_1, C_2$}{C0, C1, C2} coefficients}

            The second approach involves learning the $C_0, C_1, C_2$ coefficients of
            formula~(\ref{eq:ABC_alpha}) for the spin weight $wt$. Once learned, these coefficients 
            can be used to predict $wt$ and $wt^{norm}$ for a given $\alpha^{CP}$ hypothesis. The
            coefficients $C_0, C_1$, and $C_2$ represent physical observables, being the products of 
            longitudinal and transverse components of polarimetric vectors, as detailed in 
            formulas~(\ref{eq:CPcoeff}). 

            The classification technique using the DNN is configured to learn each $C_i$ 
            independently. The allowed ranges for these coefficients are well-established: $C_0$ 
            spans $(0.0, 2.0)$ and $C_1, C_2$ span $(-1.0, 1.0)$, as illustrated in 
            Figure~\ref{fig:c012s}. The allowed ranges for these coefficients are divided
            into $N_{class}$ bins, and each event is assigned an $N_{class}$-dimensional one-hot 
            encoded vector representing the corresponding $C_i$ value and serving as a label. In 
            this setup, a single class corresponds to a specific $C_i$ coefficient value. During 
            training, the DNN learns to associate per-event features with these class labels.
            The output is a probability distribution over the $N_{class}$ values, which is converted
            to a one-hot encoded representation, selecting the most probable class as the predicted 
            $C_i$ value. 

            \begin{figure}
                \begin{center} {
                    \includegraphics[width=4.5cm,angle=0]{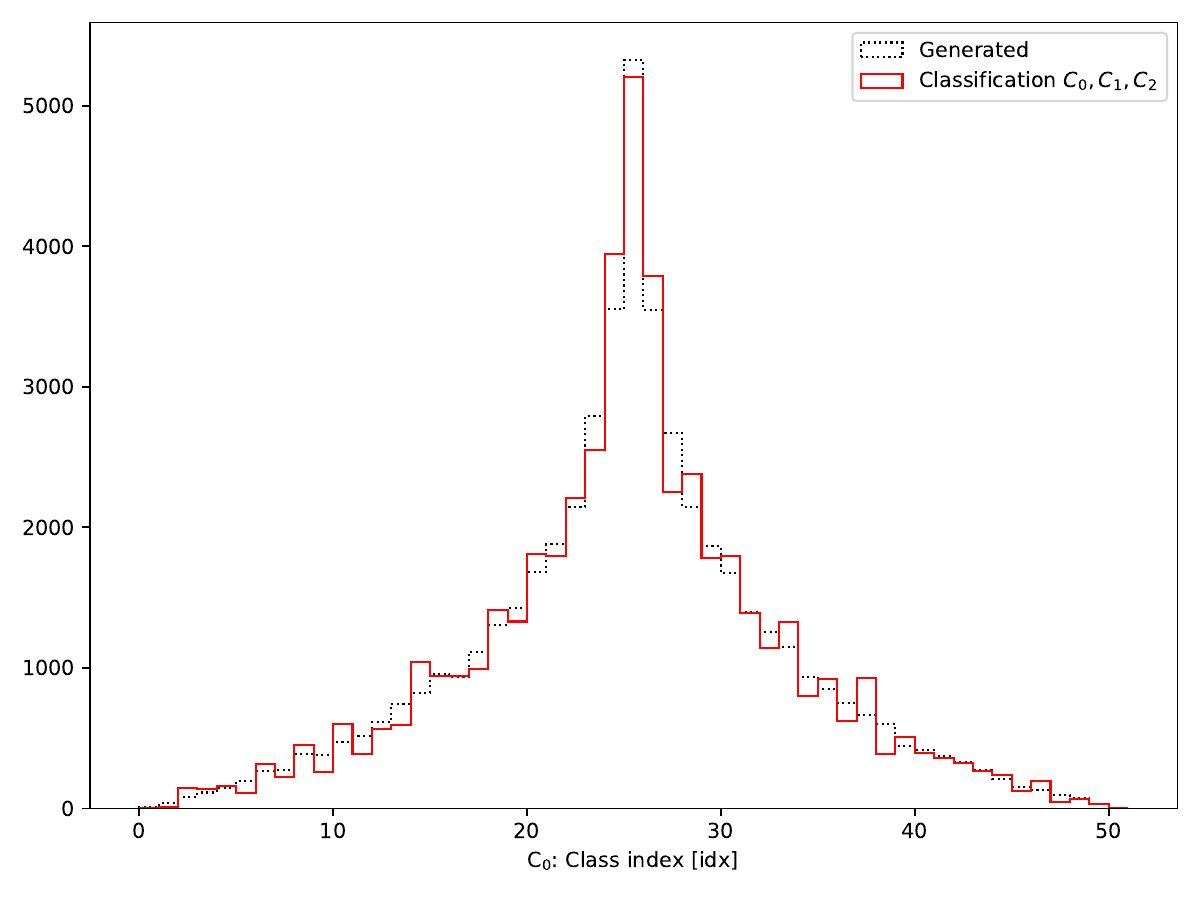}
                    \includegraphics[width=4.5cm,angle=0]{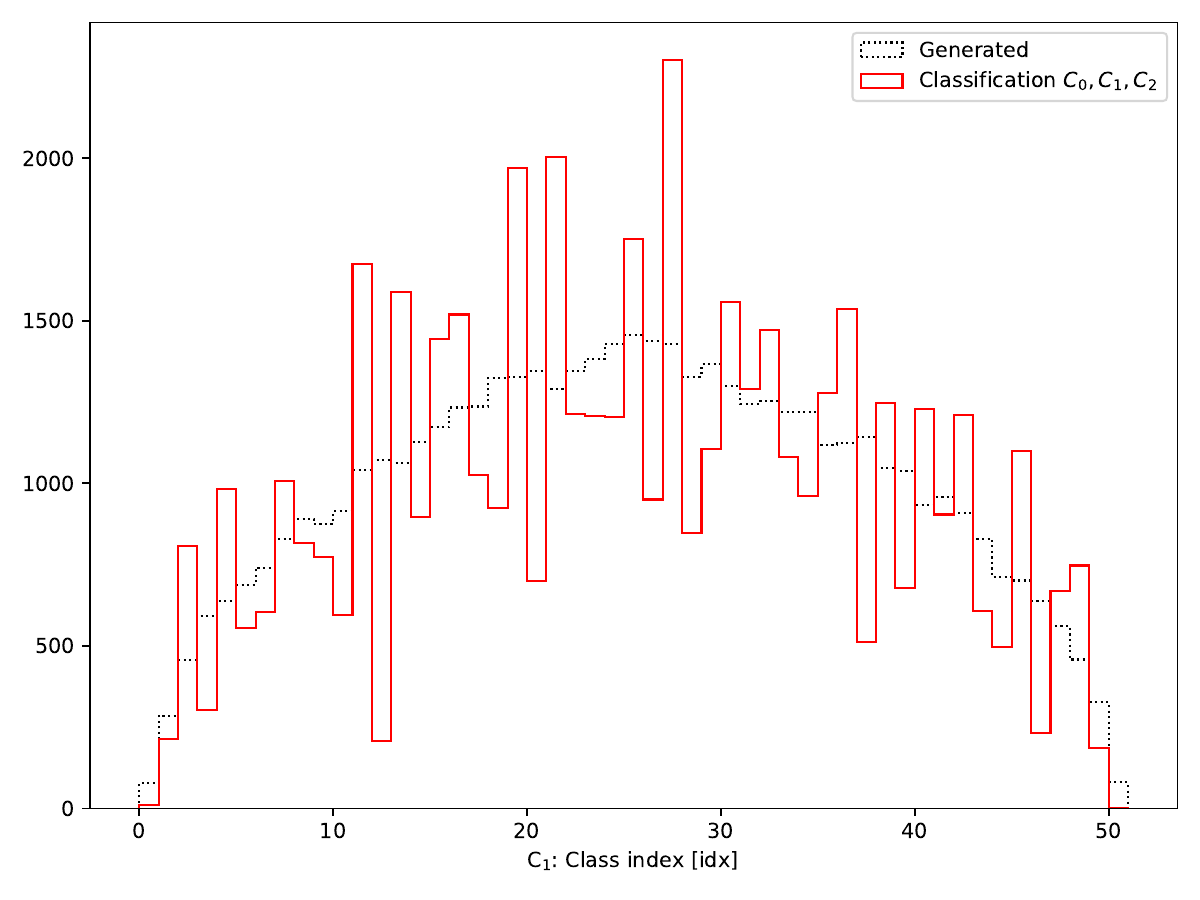}
                    \includegraphics[width=4.5cm,angle=0]{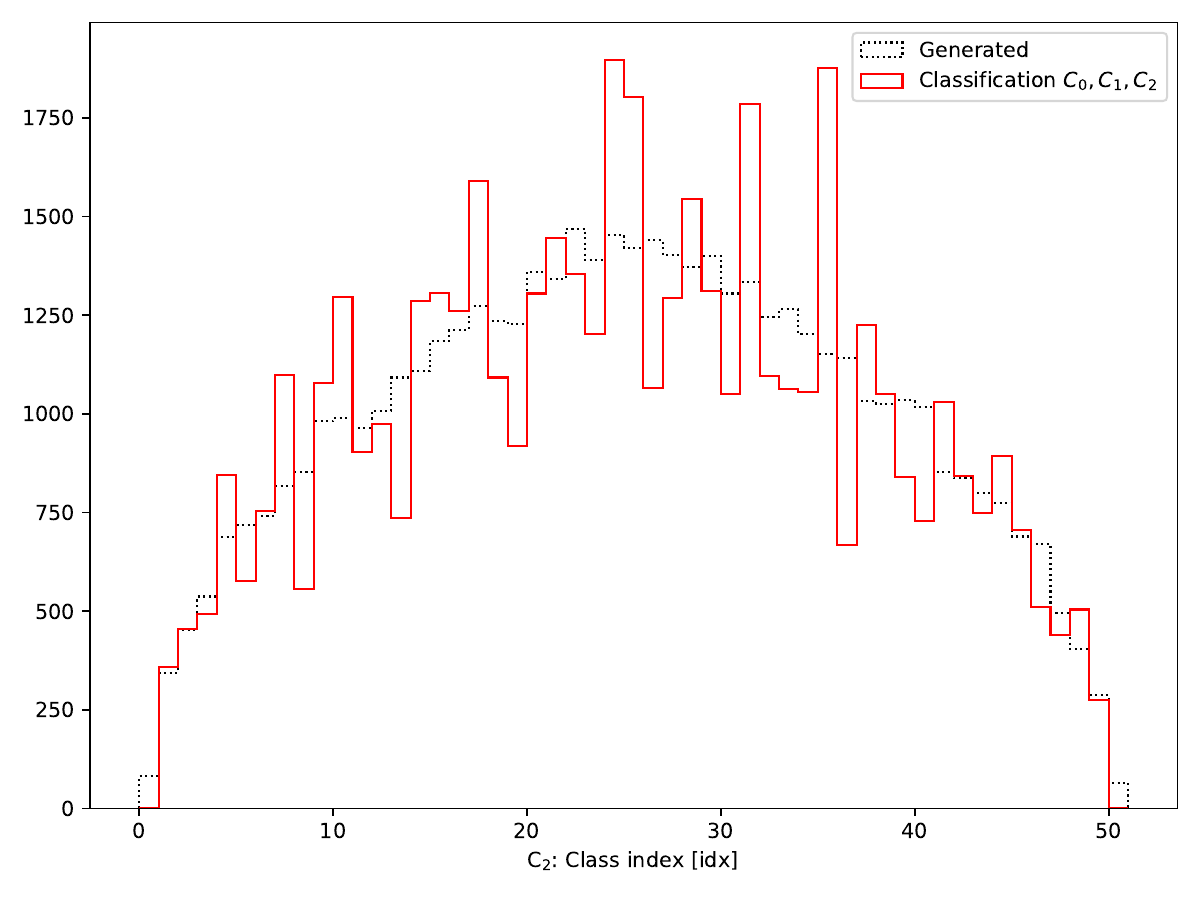}
                }
                \end{center}
                \caption{Distributions of true and predicted $C0, C1$, and $C2$ coefficients from 
                    formula~(\ref{eq:CPcoeff}). The DNN was trained using the classification method
                    with $N_{class}$ = 51 and {\it Variant-All}.}
                \label{fig:DNN_soft_nc_predC012s}
            \end{figure}

            The regression method learns all three $C_0, C_1$, and $C_2$ coefficients
            simultaneously. The learning accuracy is comparable to that of the classification method.

            We utilise the true and predicted $C_0, C_1$, and $C_2$ coefficients to compute the $wt$ 
            distribution according to formula~(\ref{eq:ABC_alpha}). This distribution is 
            then discretised into $N_{class}$ points (potentially different from the $N_{class}$ 
            used for learning coefficients), and the $\alpha^{CP}_{max}$ is determined based on the 
            class with the maximum weight. The difference between the true and predicted
            $\alpha^{CP}_{max}$ is shown in the middle row of 
            Figure~\ref{fig:DNN_delta_argmax} for $N_{class}$ = 51, comparing
            classification (left) and regression (right) methods. The Gaussian-like shape of these 
            distributions, centered around zero, clearly demonstrates successful method performance
            for {\it Variant-All}.
            The mean and standard deviation of these distributions are close to those obtained with
            the {\tt Classification:wt} approach.
        
        \subsection{Learning the \texorpdfstring{${{\alpha}^{CP}_{max}}$}{alphaCPmax}}

            The third approach is to learn the per-event most preferred mixing angle, 
            $\alpha^{CP}_{max}$, directly. For classification, the allowed range $(0, 2\pi)$ is 
            again divided into $N_{class}$ bins, where each bin represents a discrete $\alpha^{CP}$ 
            value. A one-hot encoded $N_{class}$-dimensional vector, indicating the true 
            $\alpha^{CP}$ value for each event, serves as the training label. The DNN outputs an 
            $N_{class}$-dimensional probability vector, from which the class with the highest
            probability is selected as the predicted $\alpha^{CP}_{max}$. Notably, this approach
            bypasses the prediction of the spin weight or $C_i$ coefficients. The regression method 
            implementation allows us to learn the per-event most preferred mixing angle, 
            $\alpha^{CP}_{max}$, as a continuous value without having to deal with discretised 
            predictions.
            
\begin{figure}
            \begin{center}
                {
                \includegraphics[width=8.0cm,angle=0]{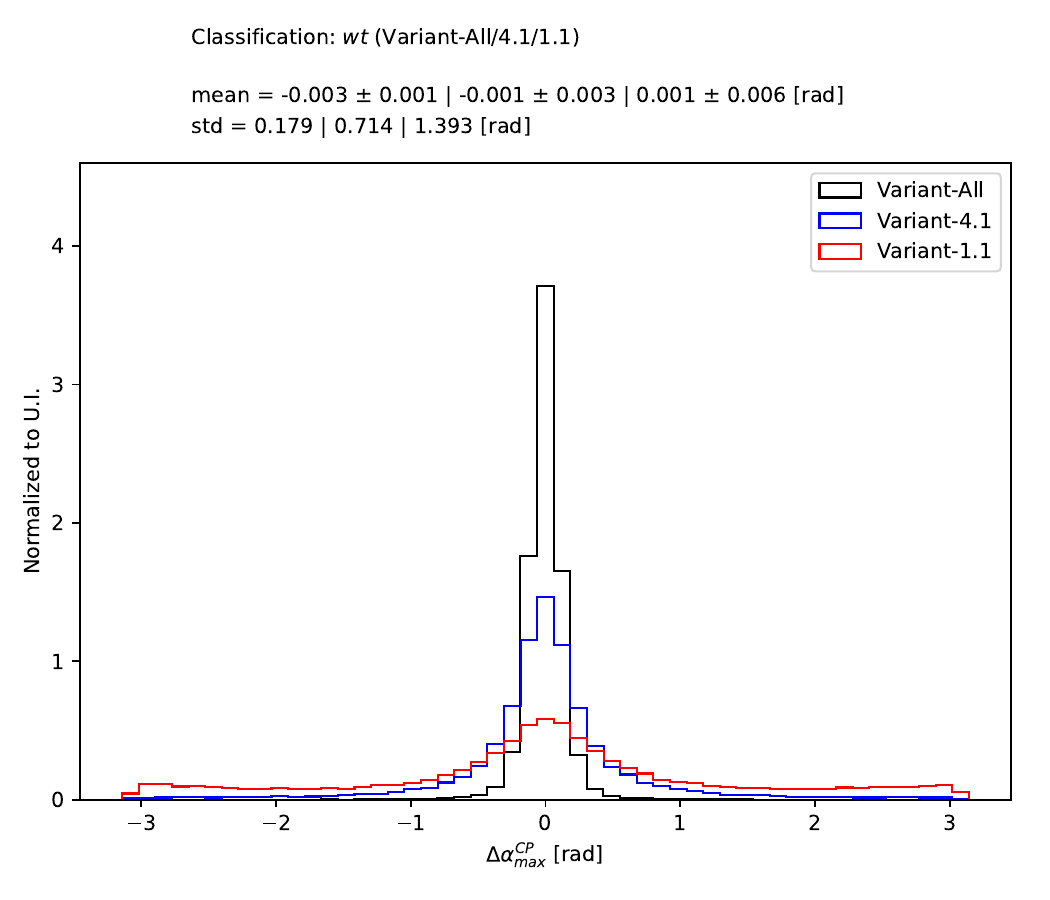}
                \includegraphics[width=8.0cm,angle=0]{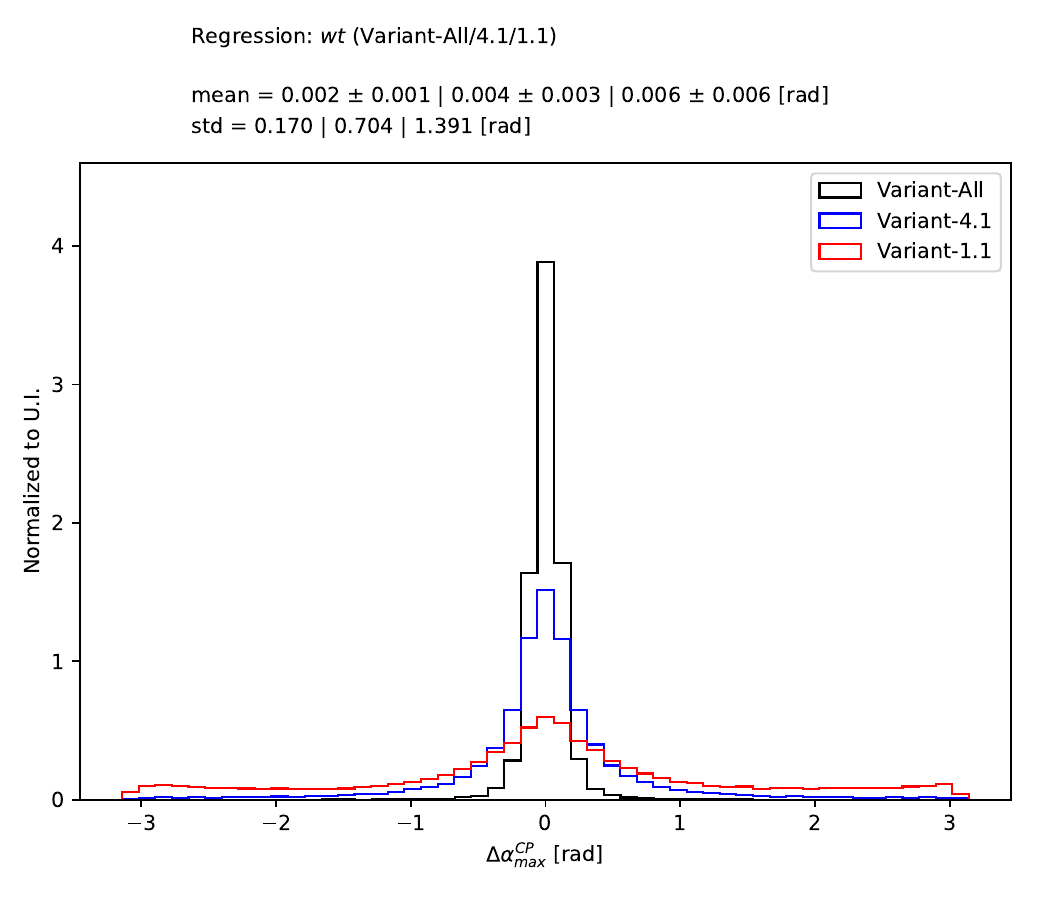}
                }
                {
                \includegraphics[width=8.0cm,angle=0]{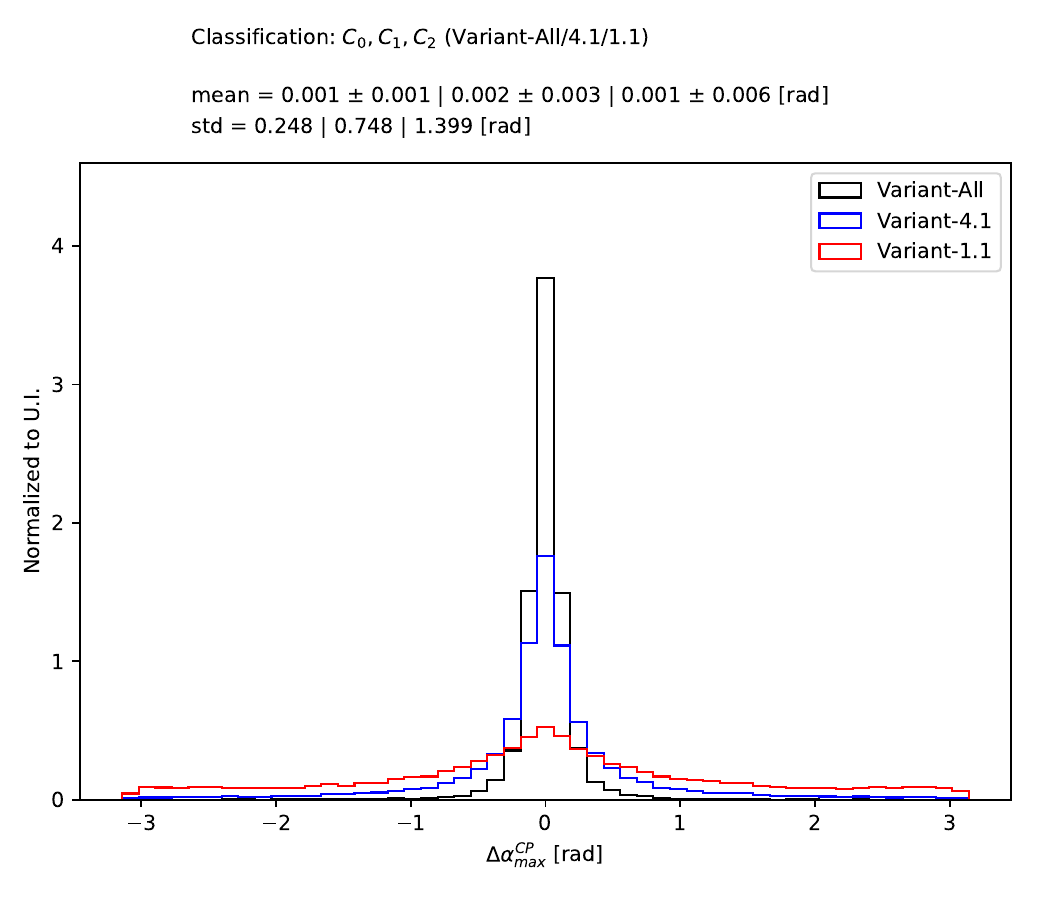}
                \includegraphics[width=8.0cm,angle=0]{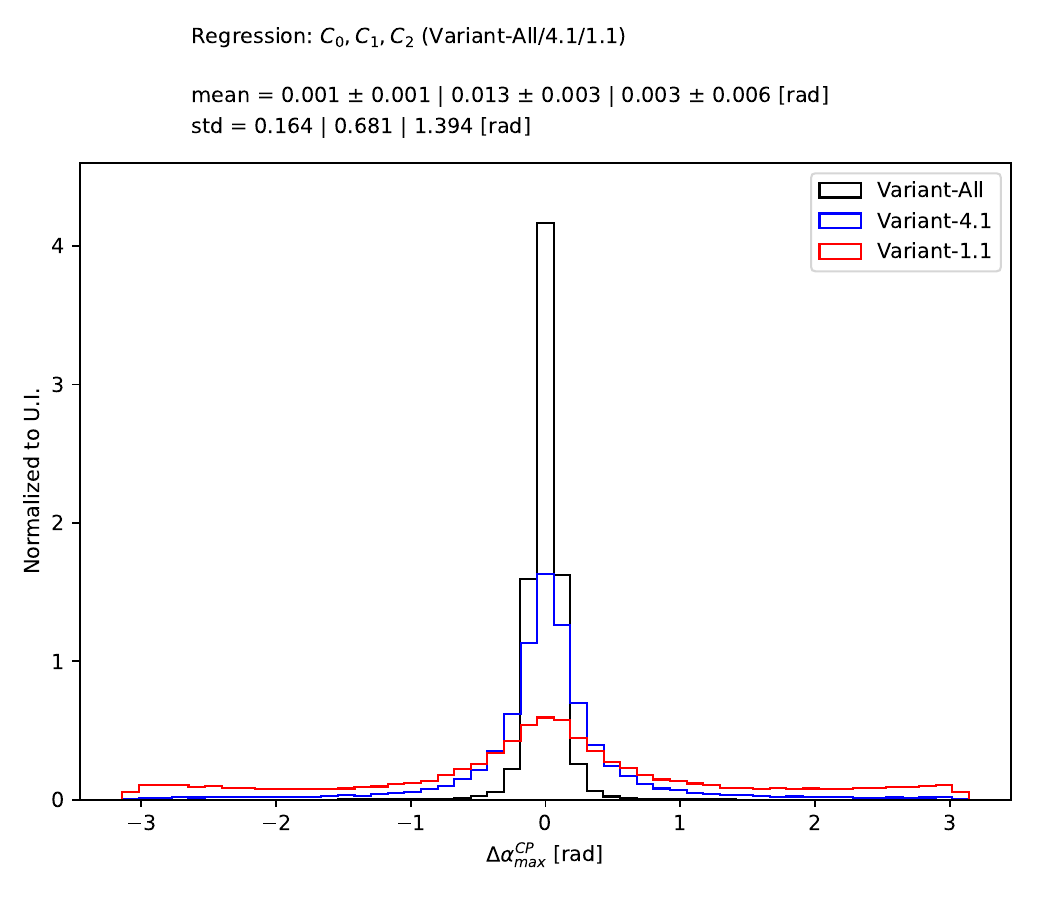}
                }
                {
                \includegraphics[width=8.0cm,angle=0]{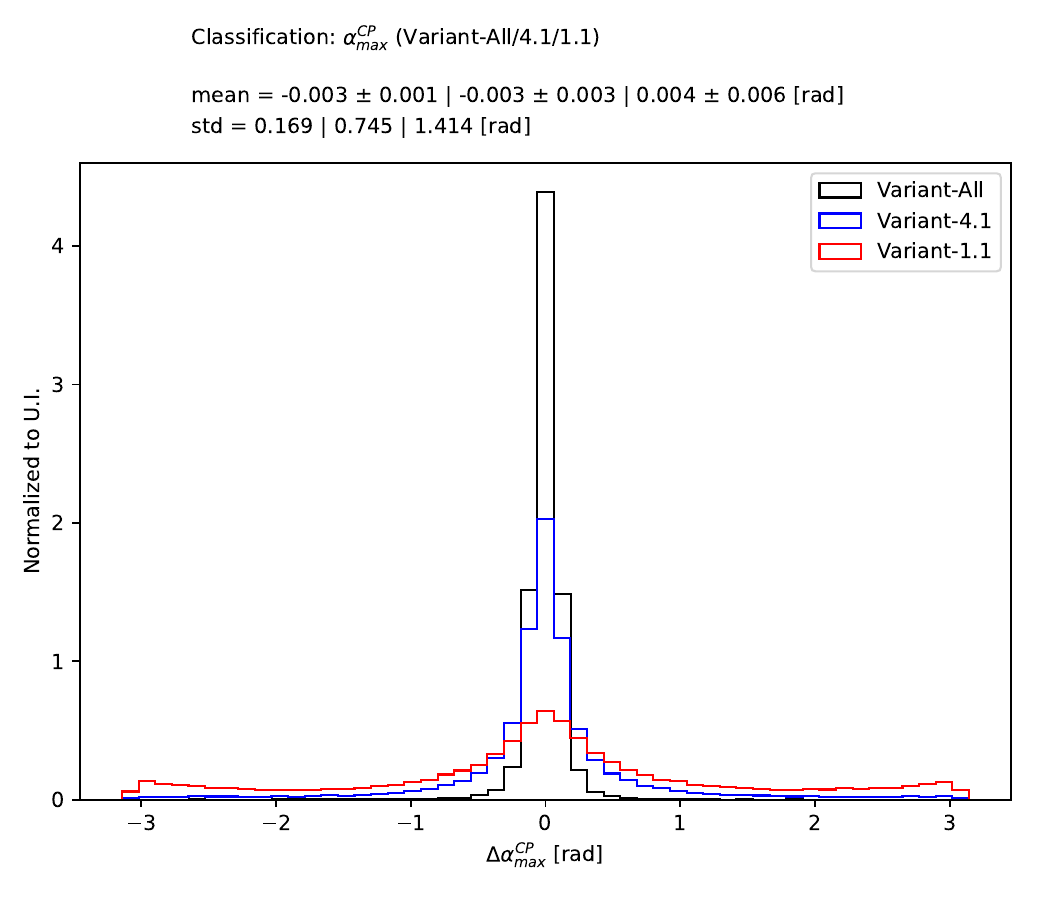} 
                \includegraphics[width=8.0cm,angle=0]{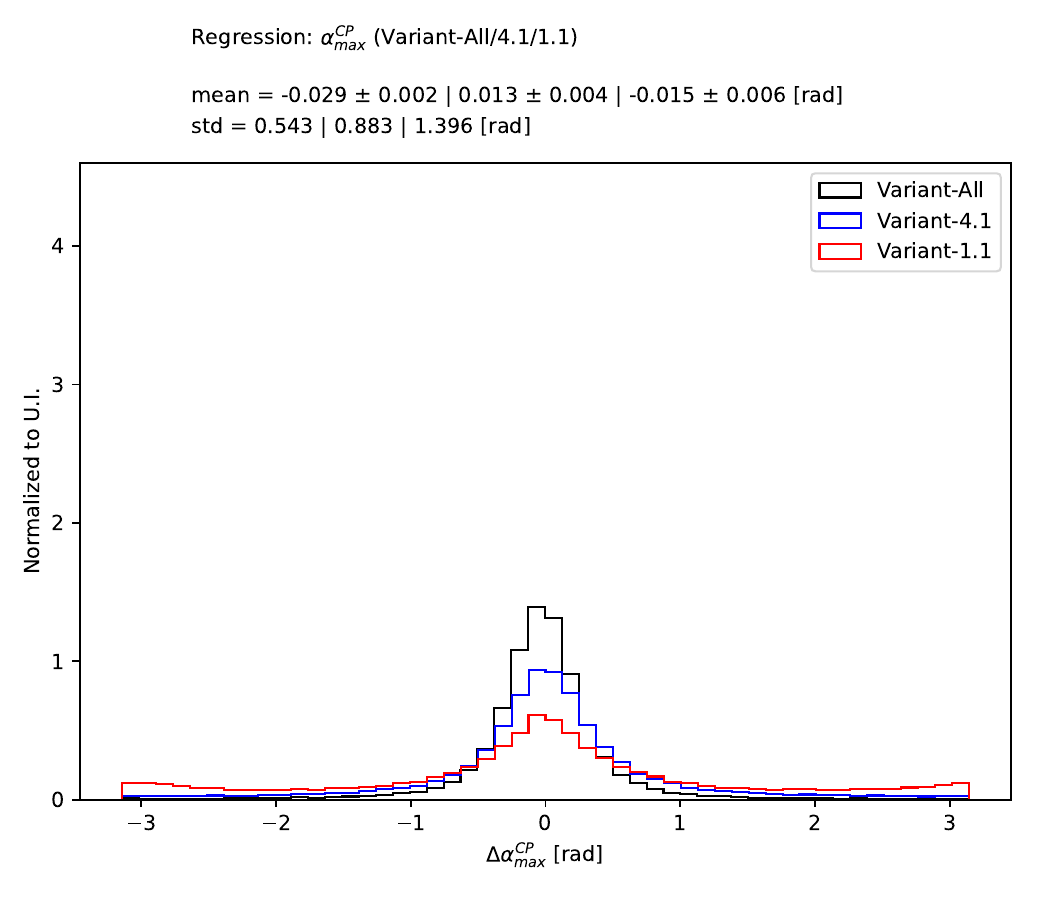}
                }
            \end{center}
            \caption{
                Distribution of the per-event difference between true and predicted 
                $\alpha^{CP}_{max}$, obtained using: (top row) spin weight learned via 
                classification (left) and regression (right); middle row: spin weight calculated 
                using formula~\ref{eq:ABC_alpha} and $C_0, C_1$, and $C_2$ coefficients
                learned via classification (left) and regression (right); bottom row: direct 
                comparison of true and predicted $\alpha^{CP}_{max}$ using classification (left) and 
                regression (right). The DNN was trained with $N_{class}$ = 51 and {\it Variant-All},
                {\it Variant-4.1}, {\it Variant-1.1}.
            }
            \label{fig:DNN_delta_argmax}
            \end{figure}

            The difference between the true and predicted $\alpha^{CP}_{max}$, denoted as 
            $\Delta \alpha^{CP}_{max}$, is shown in the bottom plot of 
            Figure~\ref{fig:DNN_delta_argmax} for both classification (left) and 
            regression (right) methods. The distribution for classification is Gaussian-like with 
            a standard deviation of 0.169~[rad] for {\it Variant-All}. Regression demonstrates worse performance.

    \section{Training and evaluation: {\it Variant-XX} }
        \label{Sec:Training-Variant-XXX}

        For {\it Variant-All}, the information accessible for DNN training is identical to that used
        for calculating weights $wt$. Rather than employing mathematical formulas, the DNN learns
        to recognise patterns in the expected $wt$ values. In contrast, for other {\it Variant-XX} 
        feature sets, while the true value for training is computed using complete event decay 
        information, the input of the DNN is restricted. Specific details can be found in 
        Table~\ref{Tab:ML_variants}.

        Table~\ref{Tab:DeltalphaCPmax-XX} compares performance on the most critical metric: the 
        difference between the true and predicted per-event spin weight $wt$ or the true and 
        predicted most preferred $\alpha^{CP}_{max}$ where the spin weight of the event peaks. 
        While the information provided to the DNN is reduced gradually from {\it Variant-4.1}
        to {\it Variant-1.1}, performance deteriorates.

        {\it Variant-1.1} offers sufficient information to calculate the $\alpha^{CP}$-sensitive 
        observable $\phi^*$ and also includes this observable explicitly, but falls short in 
        training the DNN to accurately learn variables directly tied to polarimetric vectors: the 
        spin weight $wt$ and coefficients $C_0, C_1$, and $C_2$. As shown in 
        Equation~(\ref{eq:h_master}), polarimetric vectors rely on neutrino directions which are 
        experimentally undetectable. Introducing approximations for neutrino information within 
        feature sets, as in {\it Variant-4.1}, enhances prediction accuracy.

        \begin{table}
            \vspace{2mm}
            \caption{
                Mean and standard deviation of the per-event difference ($\Delta \alpha^{CP}_{max}$)
                between true and predicted $\alpha^{CP}_{max}$ values, obtained from DNN 
                classification and regression methods for {\it Variant-All}, {\it Variant-4.1},
                and {\it Variant-1.1}.}     
                \label{Tab:DeltalphaCPmax-XX}
            \begin{center}
                \begin{tabular}{|l|c|c|}
                    \hline
                    Method  & Classification Variants: All / 4.1 / 1.1& Regression Variants: 
                    All / 4.1 / 1.1 \\
                    \hline \hline 
                    Using $wt$ & std = 0.179 / 0.714 / 1.393 [rad] & std = 0.170 / 0.704 / 1.391 
                    [rad] \\
                    \hline
                    Using $C_0, C_1, C_2$ & std = 0.248 / 0.748 / 1.399 [rad] & std = 0.164 
                    / 0.681 / 1.394 [rad] \\
                    \hline
                    Direct & std  = 0.169 / 0.745 / 1.414 [rad] & std = 0.543 / 0.883 / 1.396 
                    [rad] \\
                    \hline
                \end{tabular}
            \end{center}
        \end{table}

        \newpage
   
    \section{Results with pseudo-experiments}
        \label{Sec:Results}

        The experimental measurements published by ATLAS~\cite{ATLAS:2022akr} and 
        CMS~\cite{CMS:2021sdq} Collaborations rely on fitting the one-dimensional distribution of 
        $\phi^*$, defined as the angle between the $\tau$ decay planes. This angle is sensitive to 
        the transverse spin correlations between decaying $\tau$ leptons and consequently to the 
        CP state of the Higgs boson in $H \to \tau \tau$ decays.

        The calculation of $\phi^*$ depends on the $\tau$ lepton decay modes and was developed for 
        hadron colliders in papers~\cite{Berge:2008dr,Berge:2015nua}. It is based on the 
        four-momenta of the visible $\tau$ decay products and, in the case of the $\tau \to 
        \pi^{\pm} \nu$ decay, on the impact parameter of the decay vertex. The $\phi^*$ distribution 
        for a sample of events with a given $\alpha^{CP}$ mixing state exhibits a characteristic 
        first-order trigonometric polynomial shape, with the maximum value linked to an 
        $\alpha^{CP}$ hypothesis. Following the convention used in publications, the maximum is 
        located at $\phi^* = 90^{\circ}$ for the scalar case and at $\phi^* = 0^{\circ}, 
        180^{\circ}$ for the pseudoscalar case. The position of the maximum shifts according to the 
        linear relation $\Delta \phi^{*} = 2 \Delta \phi^{CP}$.

\begin{figure}
            \begin{center} {
                \includegraphics[width=16.0cm,angle=0]{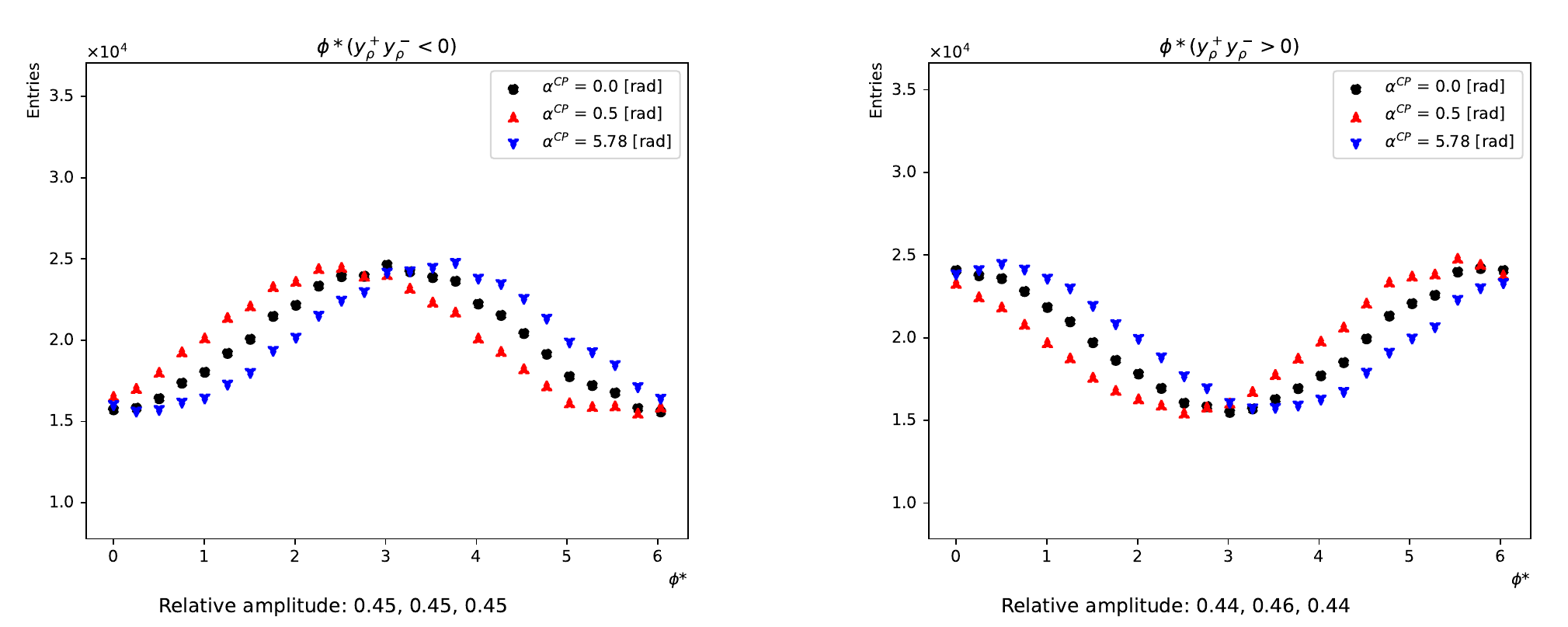}
            }
            \end{center}
            \caption{
                The $\phi^*$ distribution for three $\alpha^{CP}$ hypotheses, $\alpha^{CP} = 
                -29^{\circ}, 0^{\circ}, 29^{\circ}$.}
            \label{fig:Classic_phistar}
        \end{figure}

        Here, we employ the definition developed much 
        earlier~\cite{Bower:2002zx, Jozefowicz:2016kvz}, specifically tailored for 
        $\tau \rightarrow \rho \nu$ decays.
        Figure~\ref{fig:Classic_phistar} shows the $\phi^*$ distribution for events used in this 
        analysis for three different hypotheses: $\alpha^{CP} = -29^{\circ}, 0^{\circ}, 29^{\circ}$, 
        with the middle one corresponding to the scalar Higgs case.

        \subsection{Distribution of the spin weight \texorpdfstring{$wt$}{wt}}
            In this paper, we propose an alternative or complementary observable, the spin weight 
            $wt$ distribution, which represents the probability of an event occurring for a given 
            $\alpha^{CP}$ hypothesis. The position of the maximum of the $wt$ distribution for a 
            series of events indicates the $\alpha^{CP}$ mixing value for that series. As discussed 
            in previous sections, we can train DNN algorithms on Monte Carlo events to predict
            the per-event spin weight $wt$ for different $\alpha^{CP}$ hypotheses. Subsequently, 
            we can apply the trained algorithm to experimental data to predict the spin weight $wt$ 
            and obtain its distribution within the analysed sample. Statistical analysis of this 
            distribution (one-dimensional fit) can yield a measurement of the most probable 
            $\alpha^{CP}$ mixing value and, consequently, the CP state of the $H \to \tau \tau$
            coupling.

            For the numerical "proof of concept" presented in this paper, we used a Monte Carlo
            unweighting technique to isolate event series corresponding to specific $\alpha^{CP}$ 
            hypotheses. For each series, the trained DNN predicted the per-event spin weight
            $wt$ as a function of different $\alpha^{CP}$ hypotheses, using either classification or
            regression methods. The distribution of the summed spin weights, $\Sigma wt$, served as 
            a one-dimensional observable for measuring the most probable $\alpha^{CP}$ mixing state.

\begin{figure}
            \begin{center}
                {
                \includegraphics[width=8.0cm,angle=0]{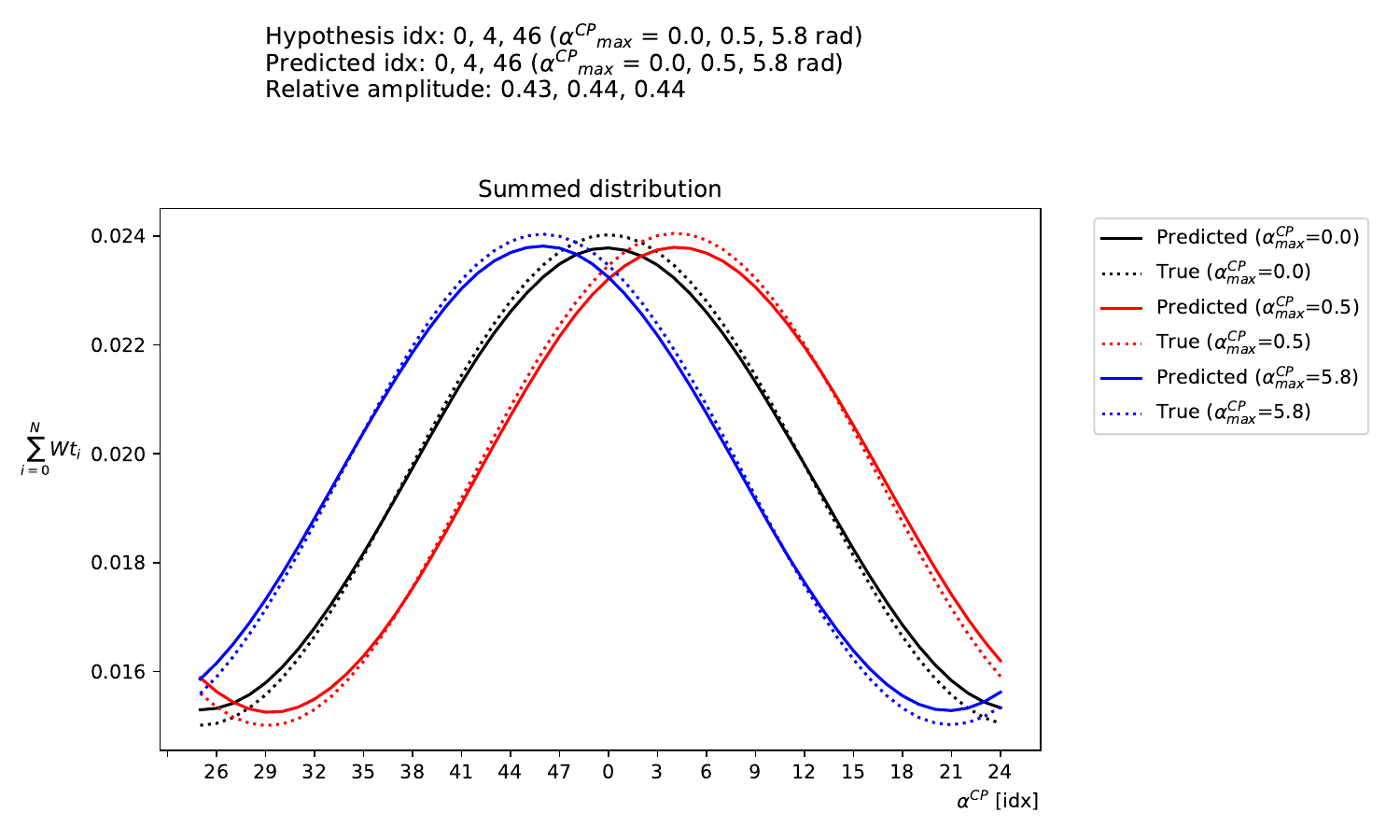}
                \includegraphics[width=8.0cm,angle=0]{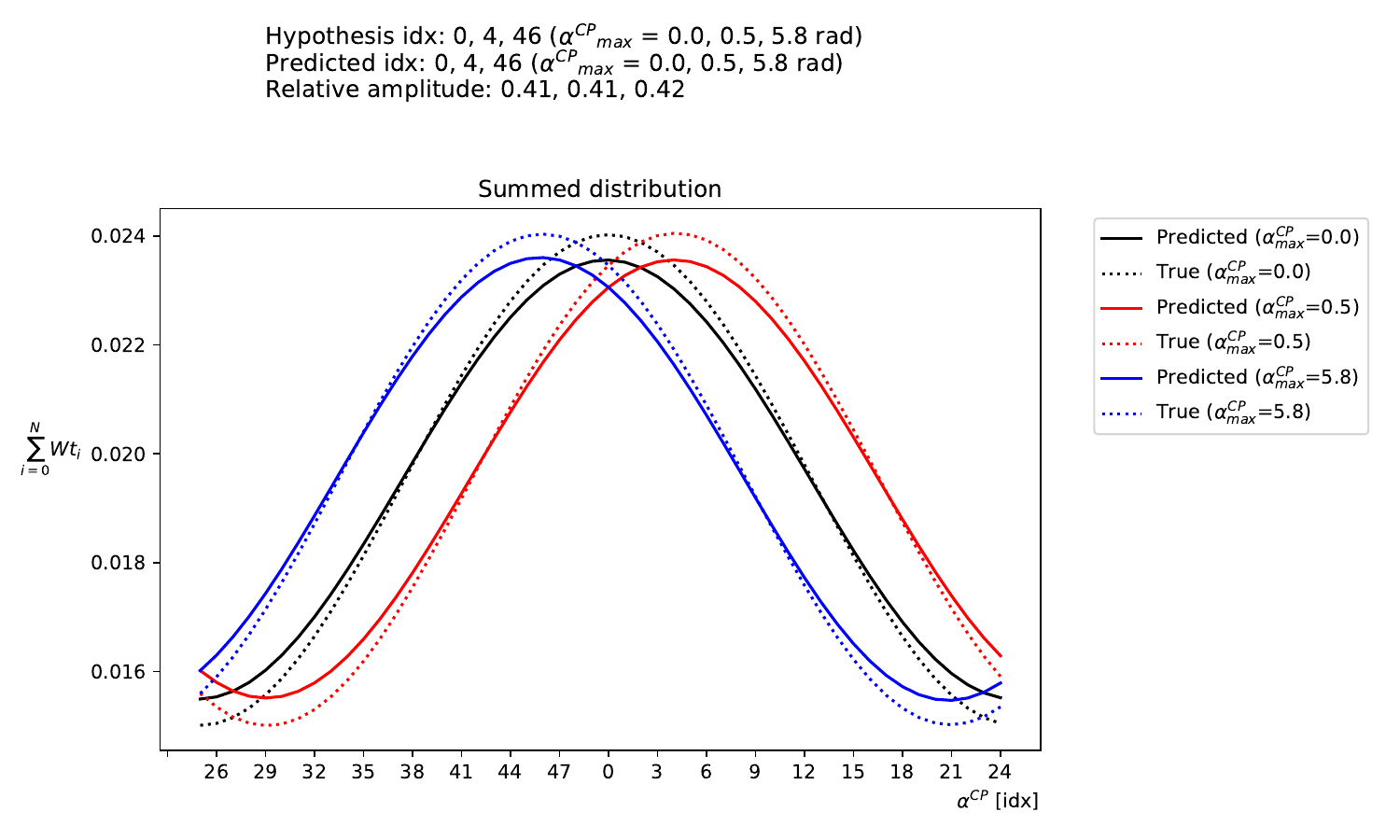}
                }
                {
                \includegraphics[width=8.0cm,angle=0]{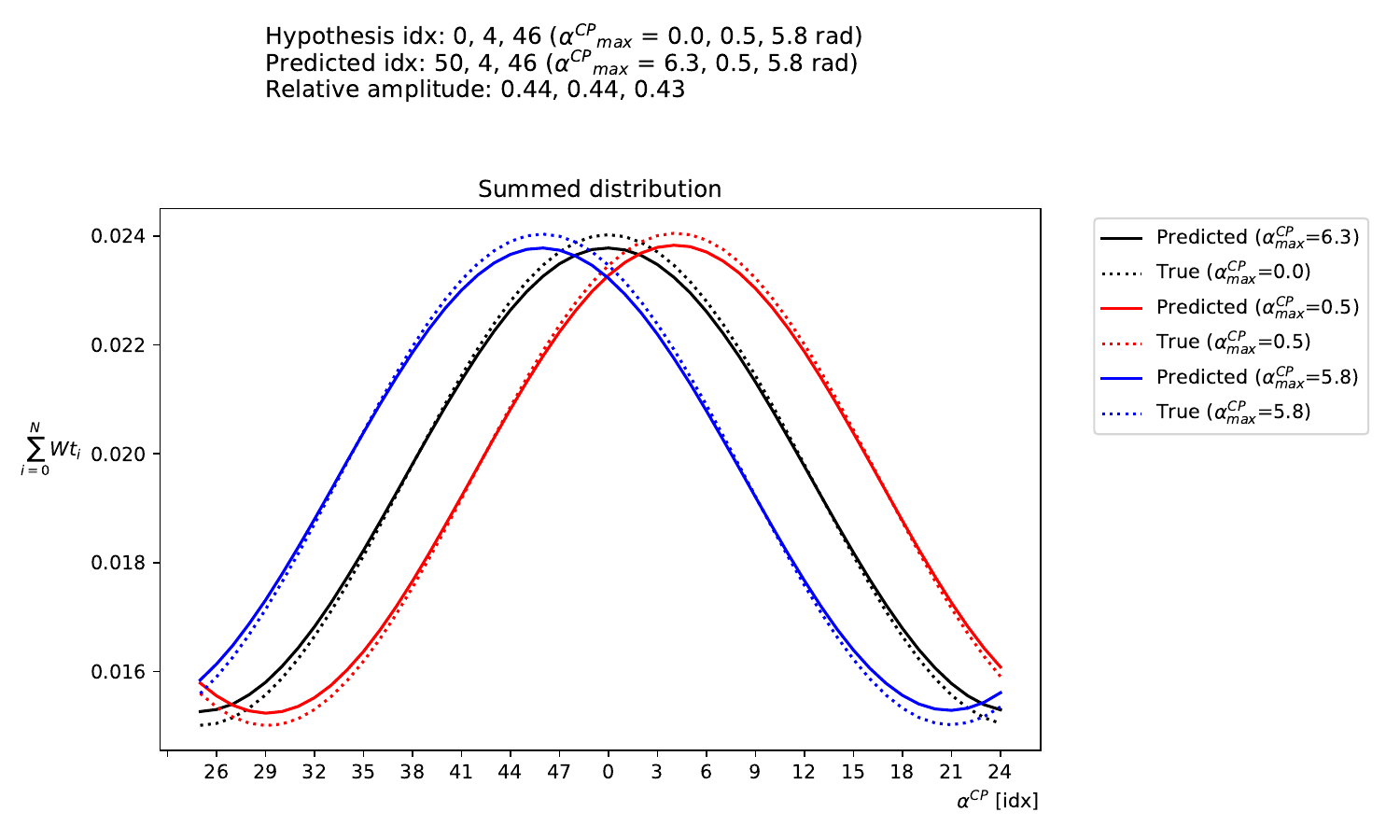}
                \includegraphics[width=8.0cm,angle=0]{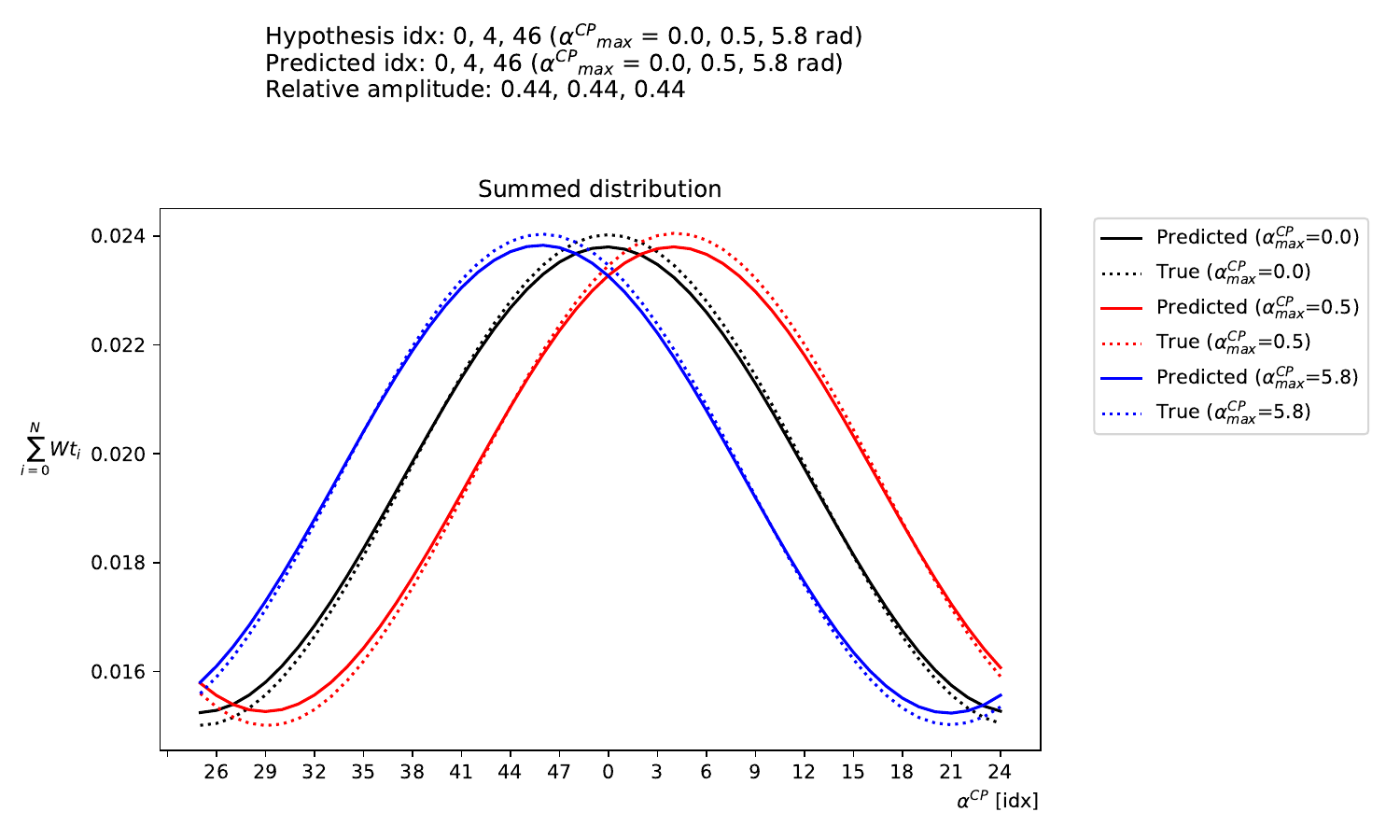}
                }
            \end{center}
            \caption{
                \raggedright Summed true and predicted per-event spin weights, $\Sigma wt$, as a 
                function of $\alpha^{CP}$ for the same series of events unweighted to $\alpha^{CP}$ 
                hypotheses with  $\alpha^{CP}= -29^{\circ}, 0^{\circ}, 29^{\circ}$. Results are shown 
                for {\tt Classification:weights} (top-left), {\tt Classification:$C_0, C_1, C_2$} 
                (top-right), {\tt Regression:weights} (bottom-left), and {\tt Regression:$C_0, C_1, 
                C_2$} (bottom-right) methods using the {\it Variant-All} feature set for DNN 
                training.
            }
            \label{fig:DNN_test_All}
        \end{figure}

\begin{figure}
            \begin{center}
                {
                \includegraphics[width=8.0cm,angle=0]{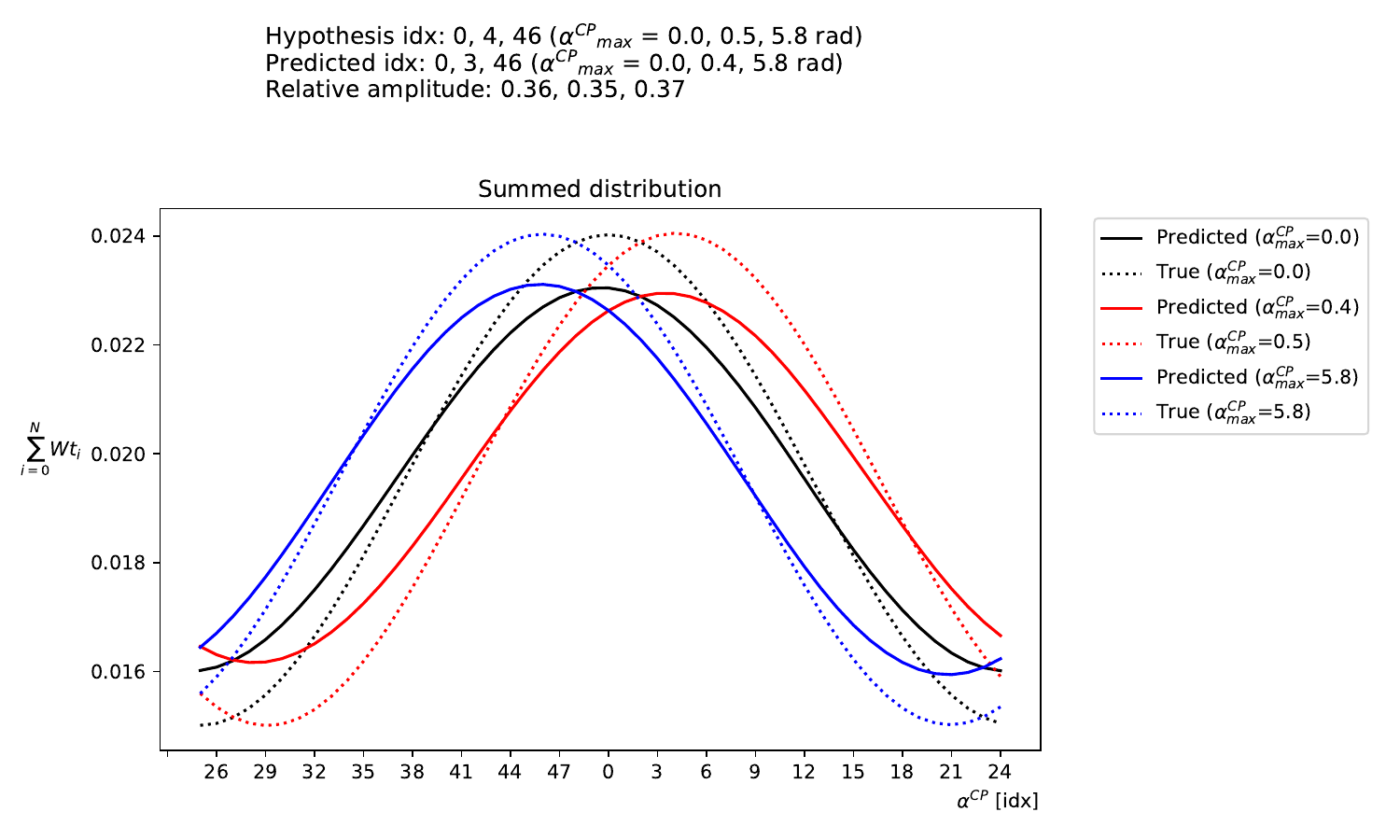}
                \includegraphics[width=8.0cm,angle=0]{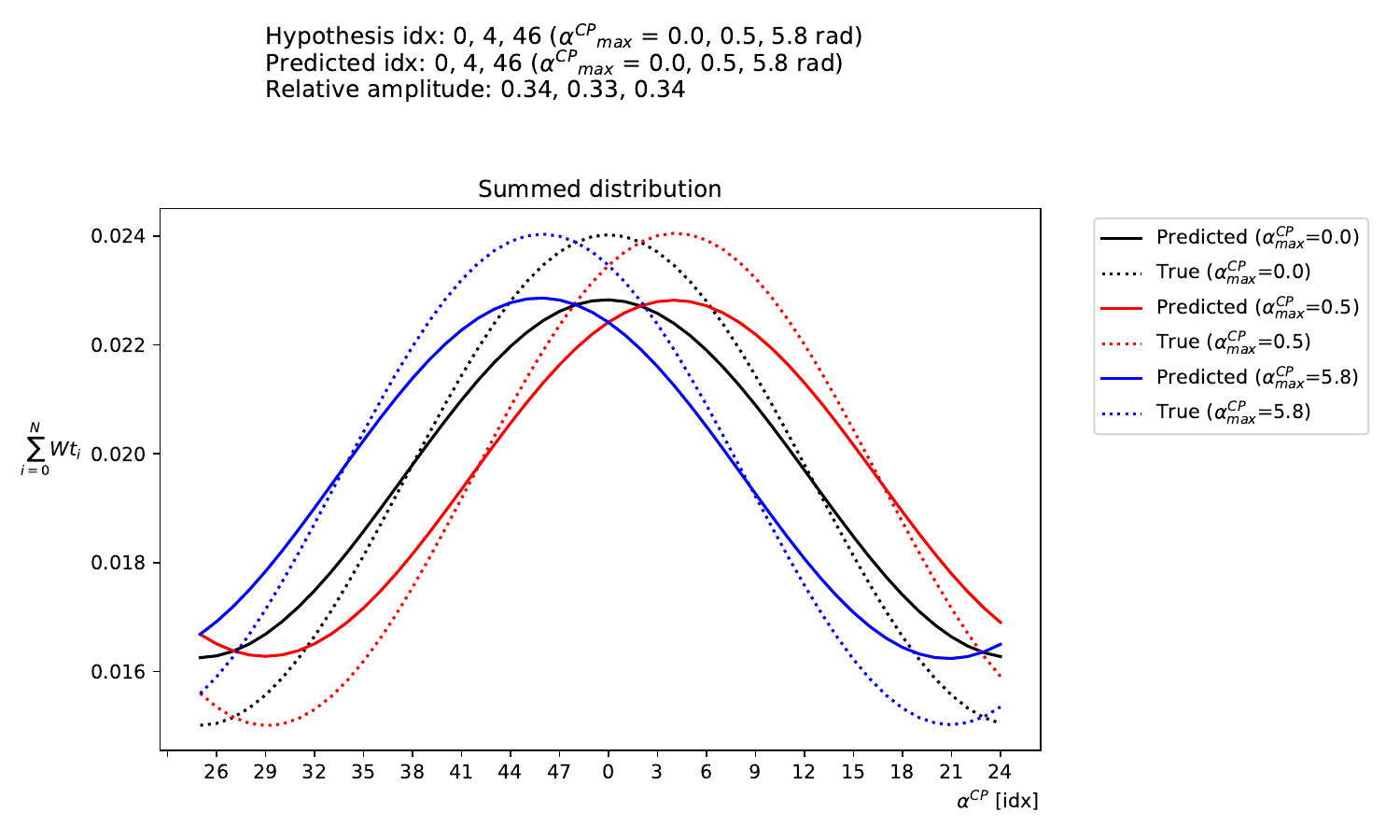}
                }
                {
                \includegraphics[width=8.0cm,angle=0]{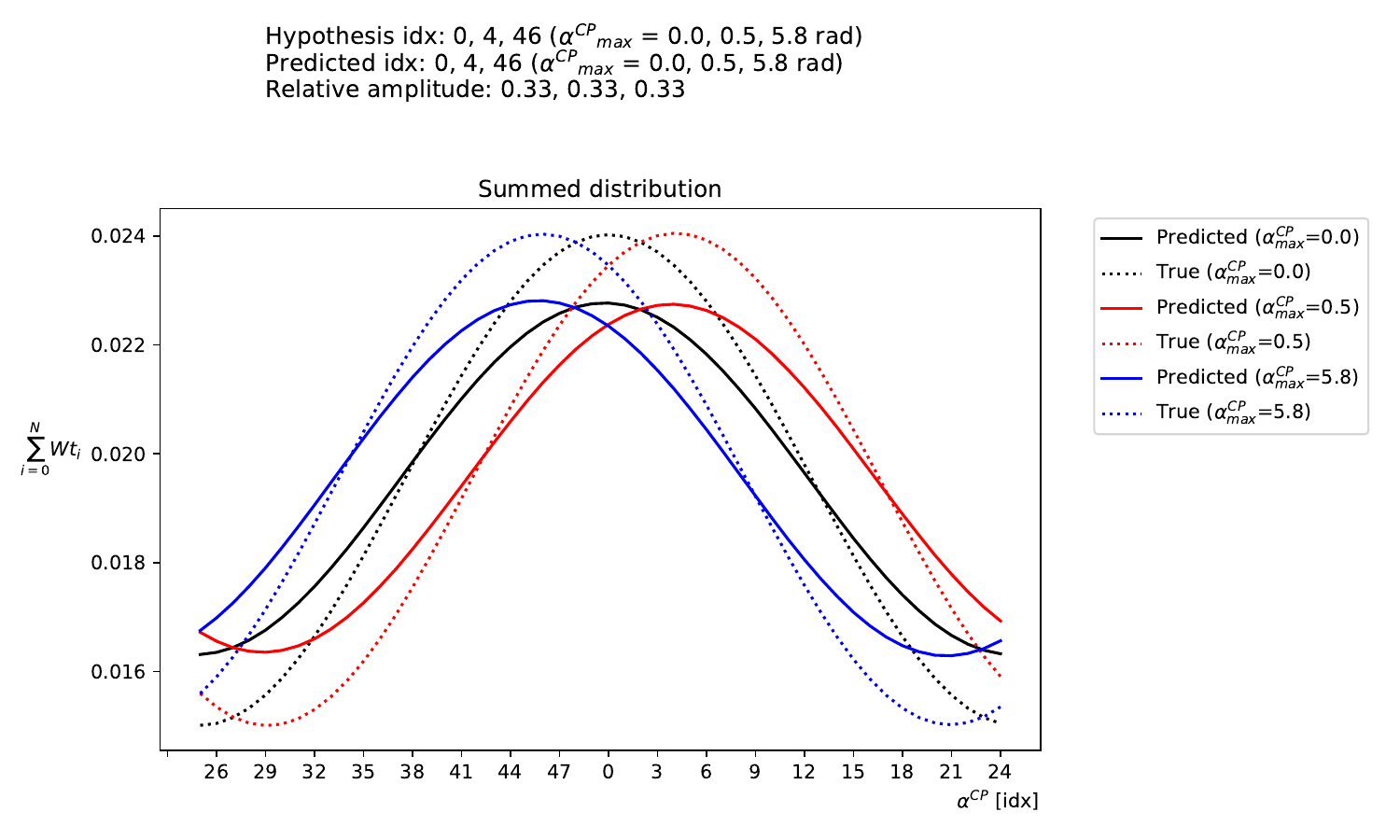}
                \includegraphics[width=8.0cm,angle=0]{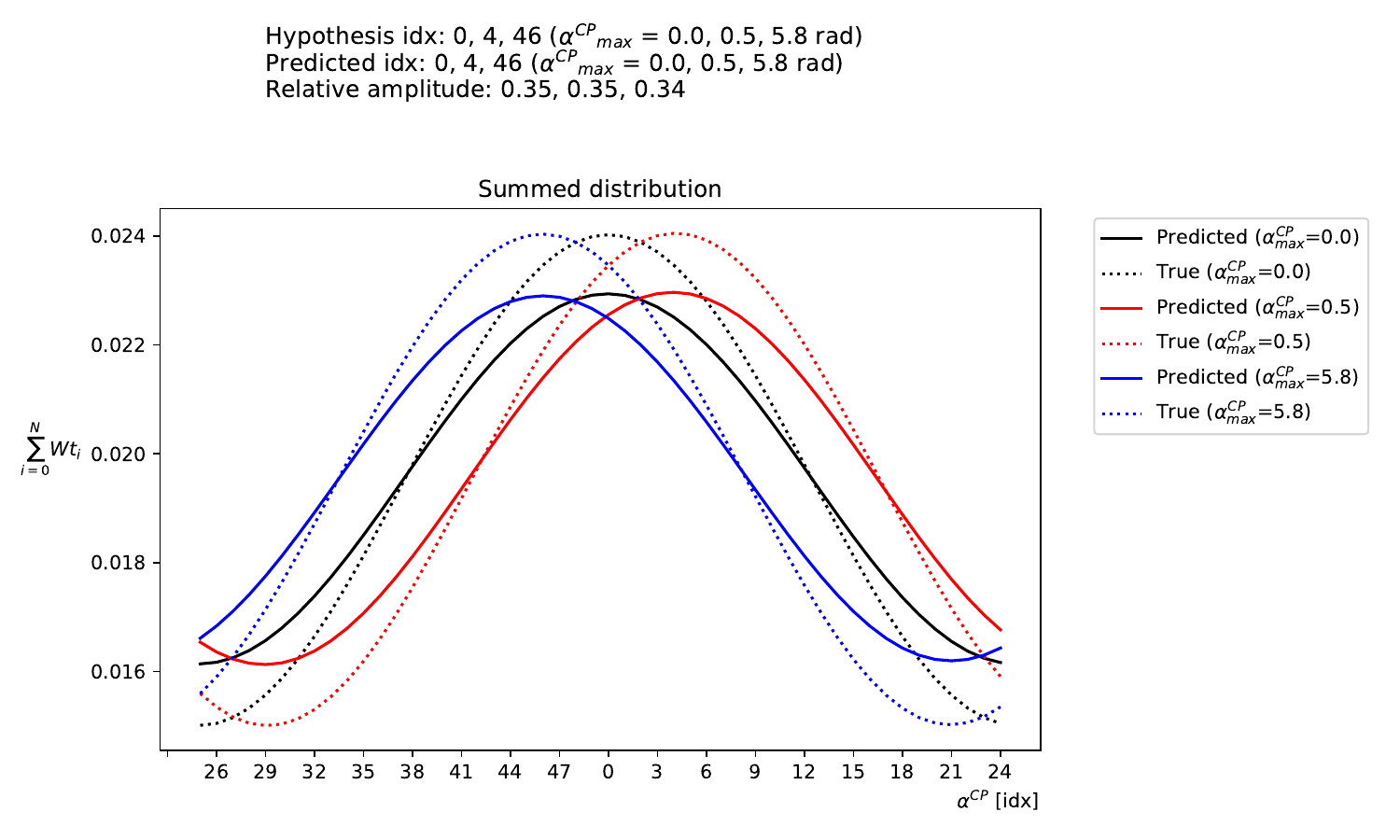}
                }
            \end{center}
            \caption{
                \raggedright Summed true and predicted per-event spin weights, $\Sigma wt$, as a 
                function of $\alpha^{CP}$ for the same series of events unweighted to $\alpha^{CP}$ 
                hypotheses with  $\alpha^{CP}= -29^{\circ}, 0^{\circ}, 29^{\circ}$. Results are shown 
                for {\tt Classification:weights} (top-left), {\tt Classification:$C_0, C_1, C_2$} 
                (top-right), {\tt Regression:weights} (bottom-left), and {\tt Regression:$C_0, C_1, 
                C_2$} (bottom-right) methods using the {\it Variant-4.1} feature set for DNN 
                training.
            }
            \label{fig:DNN_test_4.1}
        \end{figure}

\begin{figure}
            \begin{center}
                {
                \includegraphics[width=8.0cm,angle=0]{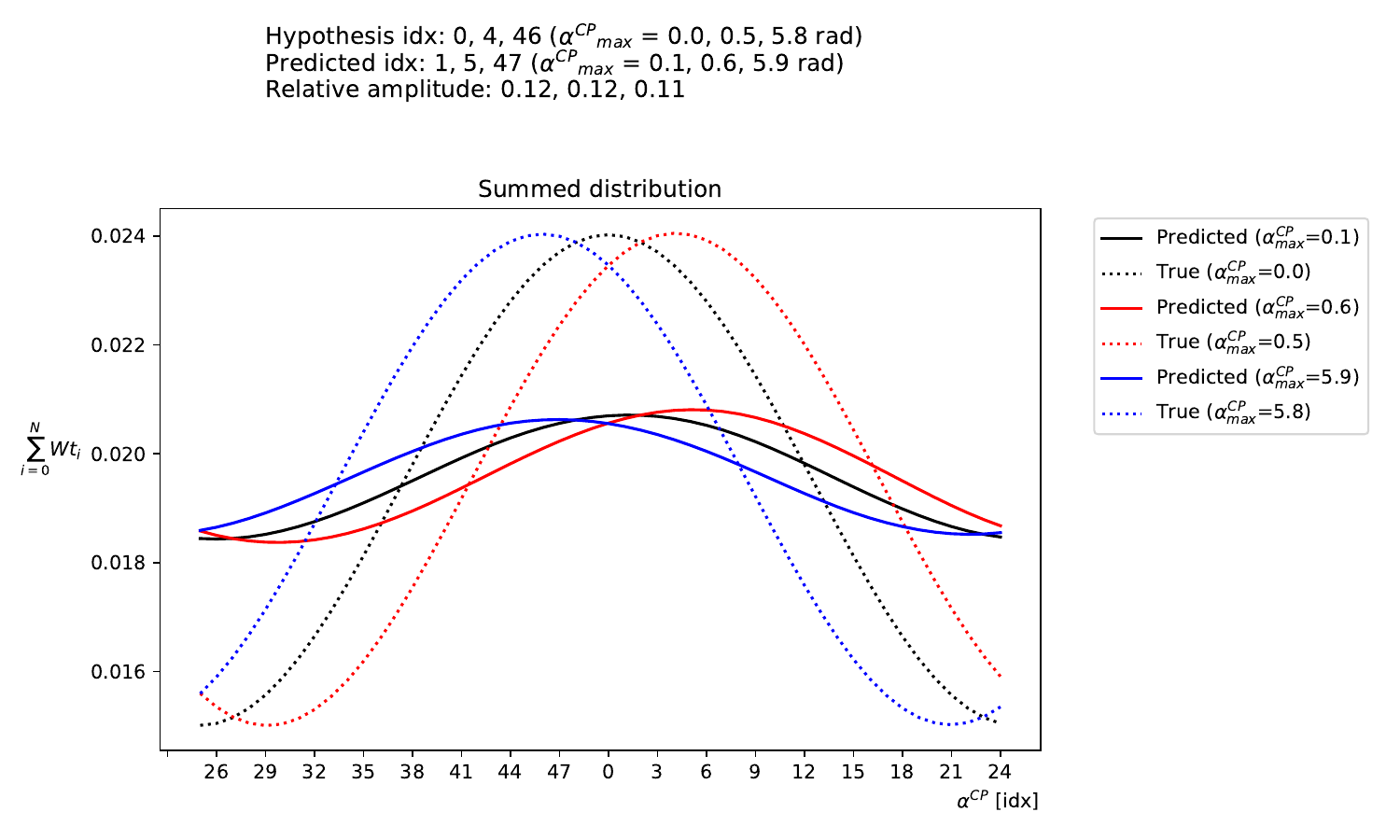}
                \includegraphics[width=8.0cm,angle=0]{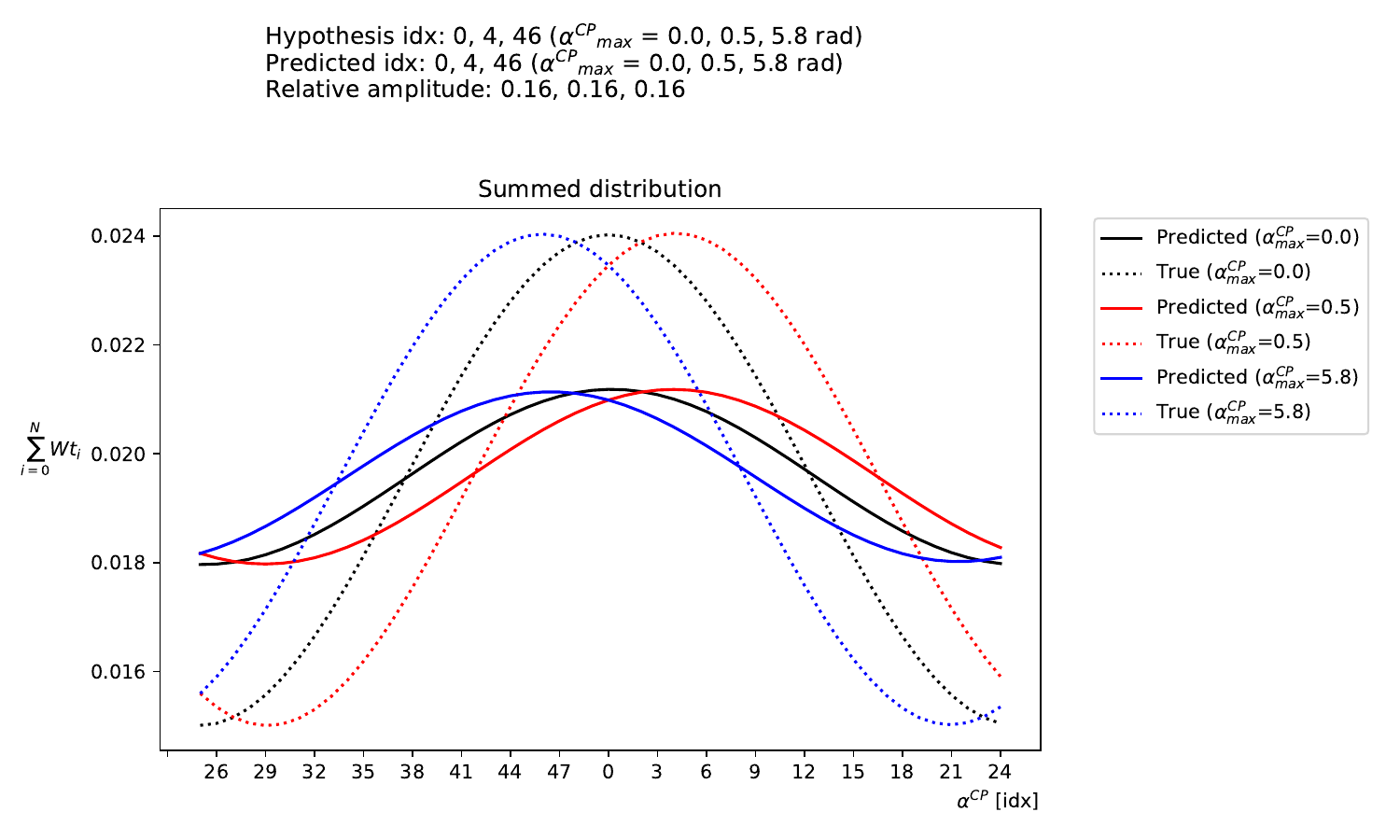}
                }
                {
                \includegraphics[width=8.0cm,angle=0]{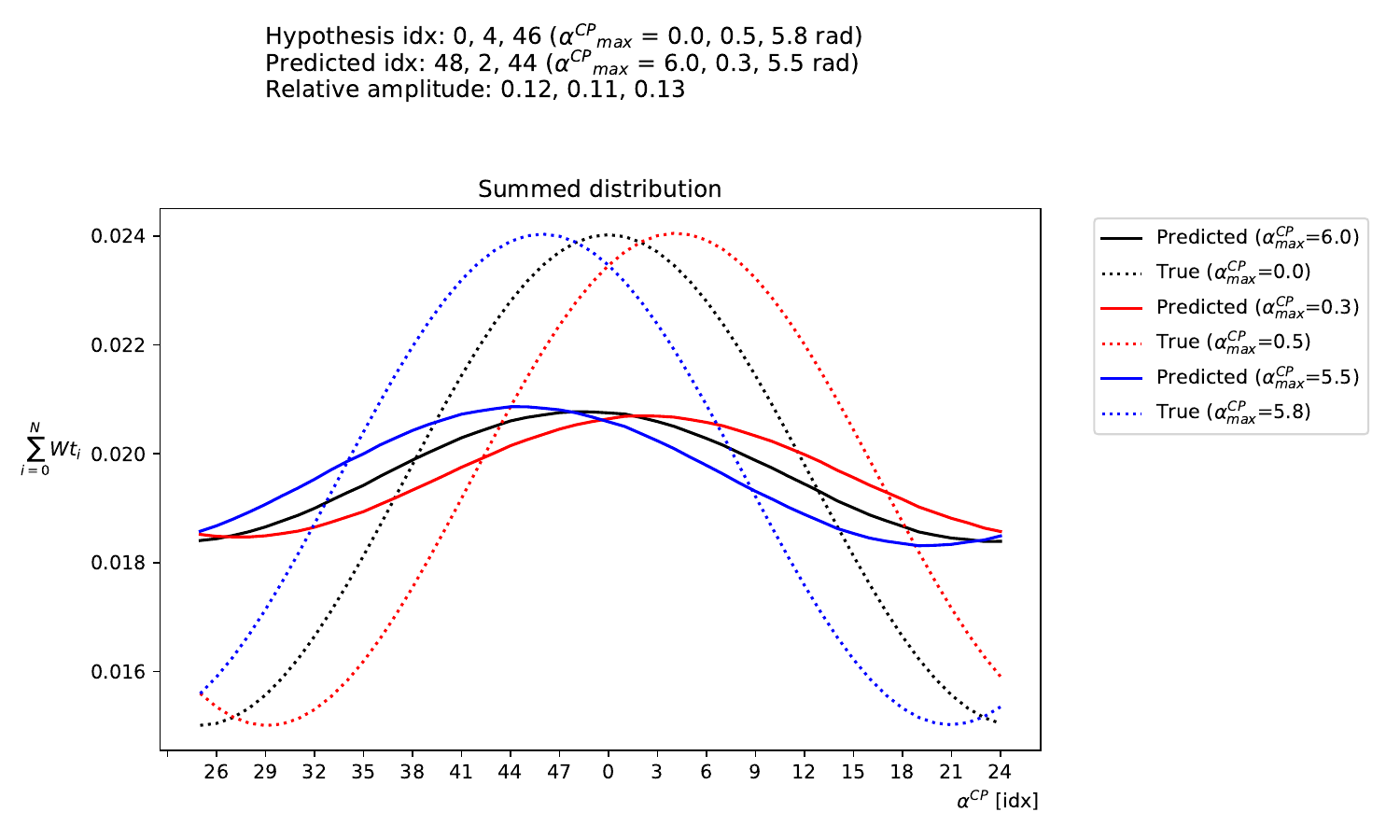}
                \includegraphics[width=8.0cm,angle=0]{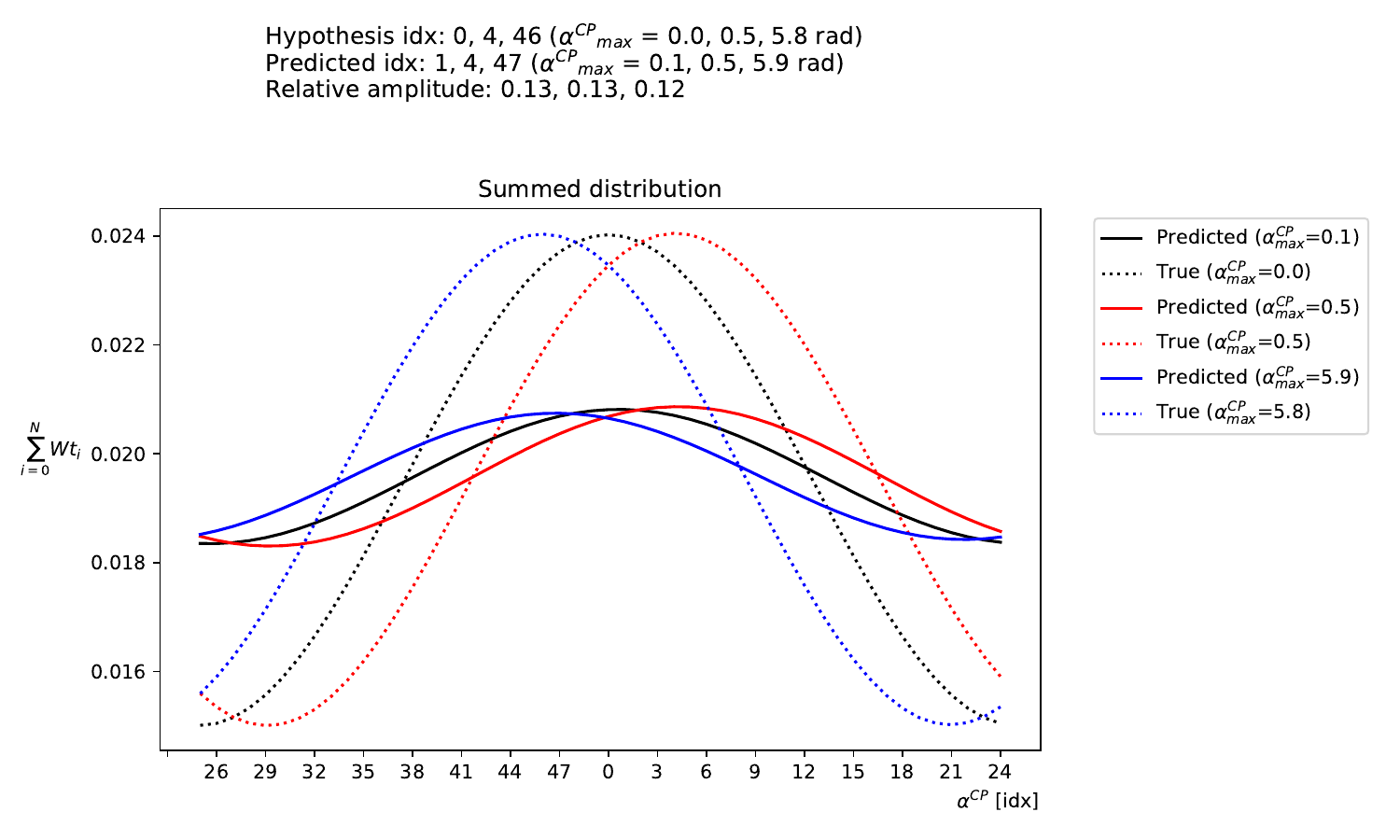}
                }
            \end{center}
            \caption{
                \raggedright Summed true and predicted per-event spin weights, $\Sigma wt$, as a 
                function of $\alpha^{CP}$ for the same series of events unweighted to $\alpha^{CP}$ 
                hypotheses with  $\alpha^{CP}= -29^{\circ}, 0^{\circ}, 29^{\circ}$. Results are shown 
                for {\tt Classification:weights} (top-left), {\tt Classification:$C_0, C_1, C_2$} 
                (top-right), {\tt Regression:weights} (bottom-left), and {\tt Regression:$C_0, C_1, 
                C_2$} (bottom-right) methods using the {\it Variant-1.1} feature set for DNN 
                training.
            }
            \label{fig:DNN_test_1.1}
        \end{figure}

            Figures~\ref{fig:DNN_test_All} to \ref{fig:DNN_test_1.1} show, for the same unweighted 
            event series corresponding to $\alpha^{CP}$ hypotheses with indices $idx = 0, 4, 46$ 
            (from the range $(0, 50)$), the sum of predicted per-event spin weights $\Sigma wt$ as
            a function of $\alpha^{CP}$ for different feature sets. While the {\it Variant-4.1} 
            feature set demonstrates relatively accurate prediction of $wt$ dependence on the 
            $\alpha^{CP}$ hypothesis, the {\it Variant-1.1} exhibits a washed-out sensitivity to 
            discriminate between hypotheses. Predicted per-event weights become flatter for most 
            events shown in the left column of Figure~\ref{fig:DNN_test_4.1}, resulting in a 
            preferred $\alpha^{CP}$ hypothesis (index $idx$ with maximum $\Sigma wt$) shifted by 
            one. This aligns with our earlier observations on training performance for different 
            feature sets in Section~\ref{Sec:Training-Variant-XXX}.

            The plots in Figures \ref{fig:DNN_test_All} to \ref{fig:DNN_test_1.1} also reveal the 
            relative amplitude of $\Sigma wt$, which diminishes as we move from {\it Variant-All} to
            {\it Variant-1.1} due to reduced information in the input feature set.

            Table~\ref{Tab:DNN_test_ampl} summarises the sensitivity of the $\Sigma wt$ observable, 
            quantified as the amplitude in the distribution as a function of the $\alpha^{CP}$
            hypothesis. Amplitude is defined as $(max-min)/((max+min)/2)$ of the $\Sigma wt$ value 
            for a series of events with a given $\alpha^{CP}$ hypothesis.

        \begin{table}
        \vspace{2mm}
            \caption{
                Amplitude of the $\Sigma wt$ distribution ($idx = 4$) obtained from DNN 
                classification and regression methods for {\it Variant-All}, {\it Variant-4.1}, 
                and {\it Variant-1.1}.
            }   
            \label{Tab:DNN_test_ampl}
            \begin{center}
            \begin{tabular}{|l|c|c|}
            \hline
            Method  & Classification Variants: All / 4.1 / 1.1& Regression Variants: 
            All / 4.1 / 1.1 \\
            \hline \hline 
            Using $wt$ & 44\% / 35\% / 12\% & 44\% / 33\% / 11\%  \\
            \hline
            Using $C_0, C_1, C_2$ & 41\% / 33\% / 16\% & 44\% / 35\% / 13\%  \\
            \hline
            \end{tabular}
        \end{center}
        \end{table}

    \section{Summary}
        \label{Sec:Summary}

        We have presented a proof-of-concept study applying DNN methods to measure the Higgs 
        boson $H \to \tau \tau$ CP-mixing angle-dependent coupling. This work extends the previous 
        research on classifying scalar and pseudoscalar Higgs CP states (Refs. 
        \cite{Jozefowicz:2016kvz, Lasocha:2018jcb}) and builds upon our earlier work on developing 
        classification and regression algorithms, in which some numerical results were collected 
        (Ref. \cite{Lasocha:2020ctd}).

        We have proposed using the per-event spin weight learned with DNN algorithms as a 
        sensitive observable for measuring the Higgs boson $H \to \tau \tau$ CP-mixing angle 
        coupling. This approach offers an alternative or complement to the commonly used $\phi^*$ 
        angle, defined as the polar angle between reconstructed $\tau$ lepton decay planes. 
        The $\phi^*$ angle has been used as a one-dimensional variable in recent 
        ATLAS~\cite{ATLAS:2022akr} and CMS~\cite{CMS:2021sdq} measurements, where a template fit was
        applied to extract the $\phi^{CP}$ value. However, it requires dedicated, per-decay mode 
        combination algorithms to reconstruct $\phi^*$ angle from the kinematics of detectable 
        $\tau$ lepton decay products. The existing algorithms might not be optimal, and their 
        development becomes even more challenging for cascade decays involving intermediate 
        resonances that decay into multi-body final states. Using the sum of predicted $wt$ 
        distribution in series of events, as a function of $\alpha_{CP}$ hypothesis seems like a 
        very interesting and promising option.

        We have extended the work of~\cite{Lasocha:2020ctd} by investigating more realistic feature
        sets for the DNN training and focusing on predicting the shape of the spin weight 
        distribution, not just the most probable $\alpha^{CP}$ mixing angle. For this 
        "proof-of-concept" study, we have considered only the dominant and most sensitive decay 
        mode,  $\tau \to \rho^{\pm} \nu$. We compare an idealised feature set, assuming complete 
        knowledge of $\tau$ decay product four-momenta including neutrinos, to more realistic 
        scenarios where only visible decay products are reconstructed or approximations are made for
        neutrinos or initial $\tau$ lepton momenta.

    \newpage
    \begin{appendices}
        \section{Neural Network}
        \label{App:DNN}
            \subsection{Architecture}
                The structure of the simulated data and the DNN architecture follows what was 
                published in our previous papers \cite{Jozefowicz:2016kvz, Lasocha:2018jcb, 
                Lasocha:2020ctd}. We use open-source libraries, TensorFlow 
                \cite{tensorflow2015-whitepaper} and Keras \cite{chollet2015keras}, to implement 
                the models, and SciPy \cite{2020SciPy-NMeth} to compute the set of 
                $C_i$ coefficients while preparing data sets.

                We consider the $H\to \tau \tau$ channel and both $\tau^{\pm} \to \rho^{\pm} \nu$ 
                decays. Each data point represents an event of Higgs boson production and $\tau$ 
                lepton pair decay products. The structure of the event is represented as 
                follows:

                \begin{equation}
                    x_i = (f_{i,1},...,f_{i,D}),w_{a_i},w_{b_i}, ...,w_{m_i}
                \end{equation}

                The  $f_{i,1},...,f_{i,D}$ represent numerical features, and 
                $w_{a_i},w_{b_i},w_{m_i} $ are weights proportional to the likelihoods that an 
                event comes from a class $A, B, ..., M$, each representing a different 
                $\alpha^{CP}$ mixing angle. The $\alpha^{CP} = 0, 2\pi$ corresponds to the 
                scalar CP state, and $\alpha^{CP} = \pi$ corresponds to the pseudoscalar CP 
                state. The weights calculated from the quantum field theory matrix elements are 
                available and stored in the simulated data files. This is a convenient 
                situation, which does not occur in many other cases of ML classification. 
                The $A, B, ... M$ distributions highly overlap in the $(f_{i,1},...,f_{i,D})$
                space.

\begin{figure}
                    \begin{center}
                        {\includegraphics[width=13.0cm,angle=0]{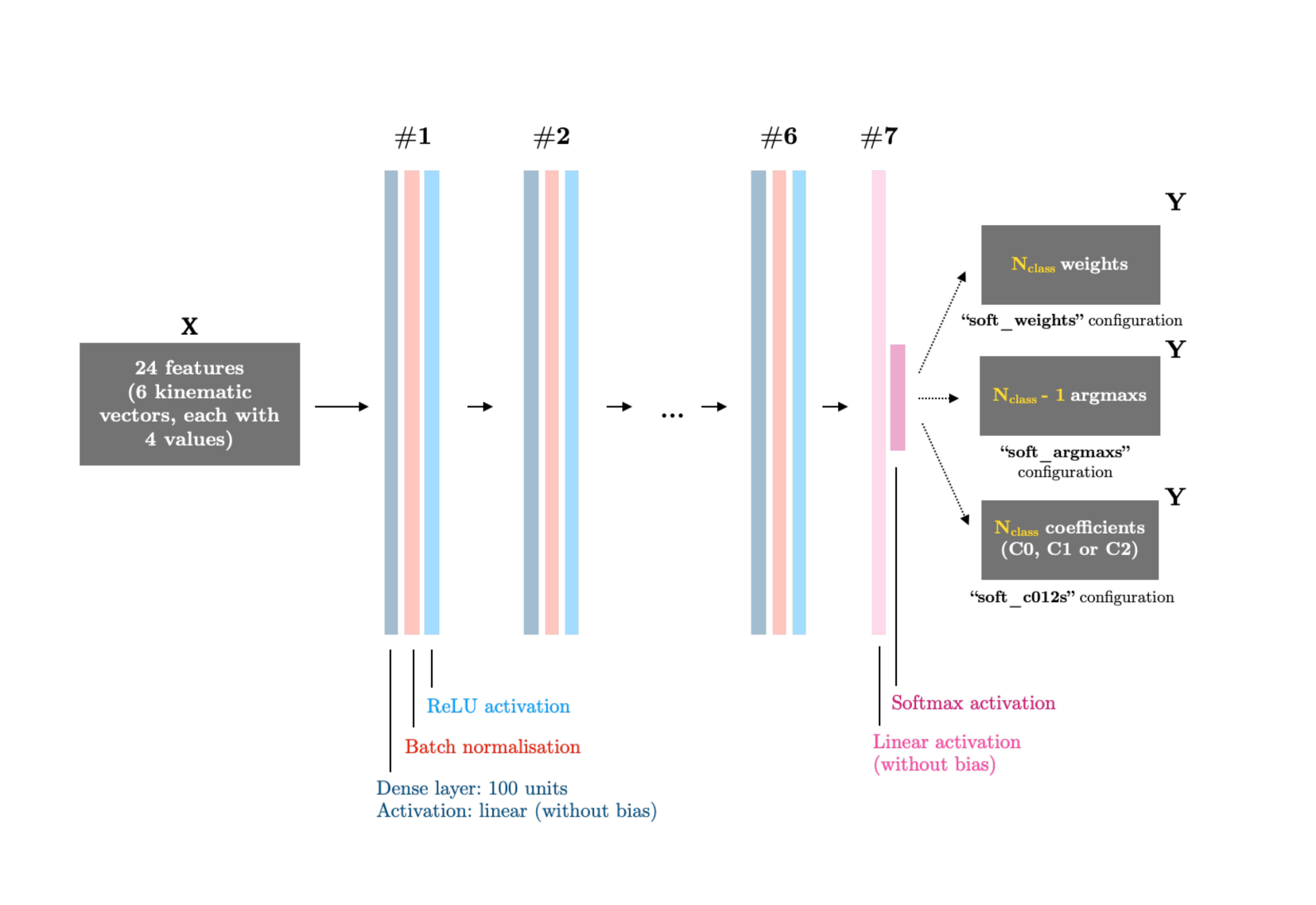}}
                    \end{center}
                    \caption{Architecture of classification models
                    \label{fig:DNN_class_architecture}}  

                    \begin{center}
                        {\includegraphics[width=13.0cm,angle=0]{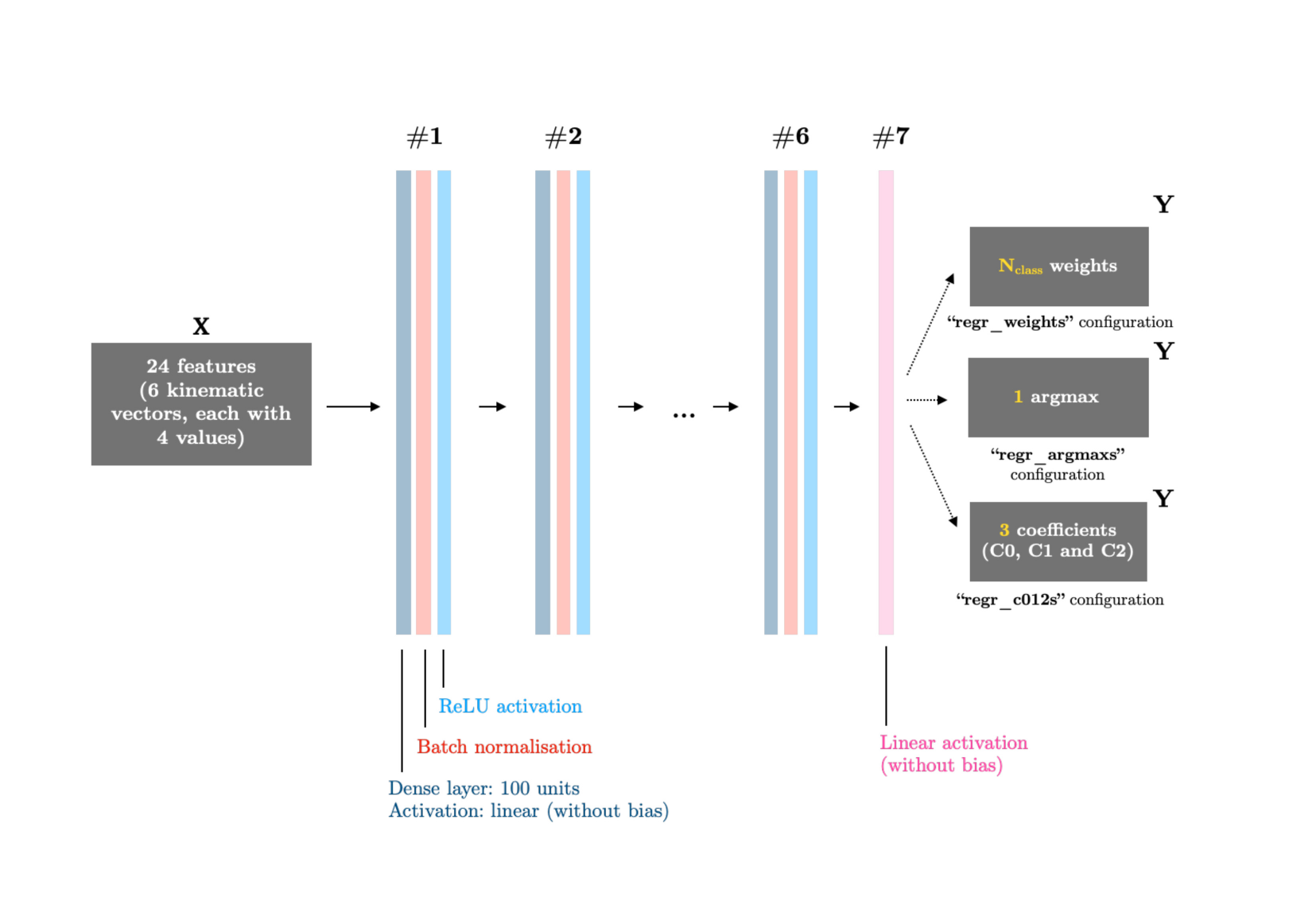}}
                    \end{center}
                    \caption{Architecture of regression models
                    \label{fig:DNN_regr_architecture}}
                \end{figure}

                Two techniques have been used to
                measure the Higgs boson CP state: multiclass classification and regression:

                \begin{itemize}
                    \item
                    For multiclass classification (Figure \ref{fig:DNN_class_architecture}), the 
                    aim is to simultaneously learn weights (probabilities) for several 
                    $\mathscr{H}_{\alpha^{CP}}$ hypotheses, learn coefficients of the weight 
                    functional form, or directly learn the mixing angle at which the spin weight 
                    has its maximum, $\alpha^{CP}_{max}$. A single class can 
                    be either a single discretised $\alpha^{CP}$ or $C_i$ coefficient value. 
                    The system learns the probabilities for classes to be associated with the event. 
                    \item
                    For the regression case (Figure \ref{fig:DNN_regr_architecture}), the aim is 
                    similar to the multiclass classification case,
                    but now the problem is defined as a continuous case. The system learns a value 
                    to be associated with the event. The value can be a vector of spin weights for a 
                    set of $\mathscr{H}_{\alpha^{CP}}$ hypotheses, a set of $C_i$ coefficients or 
                    $\alpha^{CP}_{max}$. 
                \end{itemize}

                The network architecture consists of 6 hidden layers, each with 100 nodes,
                undergoing batch normalisation~\cite{batchNorm}, followed by the ReLU activation 
                function. Model weights are initialised randomly. We use 
                Adam \cite{adamOptimiser} as an optimiser.

                The last layer is specific to the implementation case, differing in dimension of
                the output vector, activation function, and loss function. The details are
                desribed below.

                {\bf Classification:}
                The loss function used in stochastic gradient descent is the cross-entropy of 
                valid values and neural network predictions. The loss function for a sample of 
                $N_{evt}$ events and classification for $N_{class}$ classes reads as follows:

                \begin{equation}
                    Loss = - \sum_{k=1}^{N_{evt}} \sum_{i=1}^{N_{class}} y_{i,k} ln(p_{i,k}),
                \end{equation}

                where $k$ stands for consecutive events and $i$ for the class index. The 
                $p_{i,k}$ represents the neural network predicted probability for event $k$ 
                being of class $i$, while $y_{i,k}$ represents the true probability used in 
                supervised training.

                {\bf Regression:}
                In the case of predicting $wt$, the last layer is $N$-dimensional output
                (the granularity with which we want to discretise it). For predicting 
                $C_0, C_1, C_2$, the last layer is $N=3$ dimensional output, i.e. values of
                $C_0, C_1, C_2$. Activation of this layer is a linear function. The loss function 
                is defined as the Mean Squared Error (MSE) between true and predicted parameters:

                \begin{equation}
                    Loss = \sum_{k=1}^{N_{evt}} \sum_{i=1}^{i=N} (y_{i,k} - p_{i,k})^2,
                \end{equation}

                where $k$ stands for the event index and $i$ for the index of the function form 
                parameter. The $p_{i,k}$ represents the predicted value of the $C_i-th$ 
                parameter for event $k$ while $y_{i,k}$ represents the true value.

                For predicting $\alpha^{CP}_{max}$, the last layer is $N=1$ dimensional 
                output, i.e. values of $\alpha^{CP}_{max}$, with the loss function: 

                \begin{equation}
                    Loss = \sum_{k=1}^{N_{evt}} (1 - cos(y_{k} - p_{k})),
                \end{equation}

                where $p_{k}$ and $y_{k}$ denote the predicted and true value of 
                $\alpha^{CP}_{max}$, respectively. 

            \subsection{Training}
                During the training process, we monitored accuracy, the difference
                between predicted and true values (mean, $L_1$ and $L_2$), and training epoch 
                relative loss function value in the case of the multiclass classification model 
                predicting the $\alpha^{CP}_{max}$ probability distribution. For other models, 
                training and validation loss function values were recorded.

                All DNN configurations (3 multiclass classification models and 3 regression 
                models) have been trained for 25 epochs. Extending the training process to up 
                to 120 epochs did not lead to significant changes in the performance of the 
                models. During the training, model weights were saved for each epoch separately, 
                and after the training, we used the "best" weights for evaluation. The "best" 
                weights were those leading to the highest validation accuracy for the 
                multiclass classification model predicting the $\alpha^{CP}_{max}$ probability 
                distribution or the lowest loss function value for all the other DNN 
                configurations.

\begin{figure}
                    \begin{center}
                    {
                    \includegraphics[width=7.5cm,angle=0]{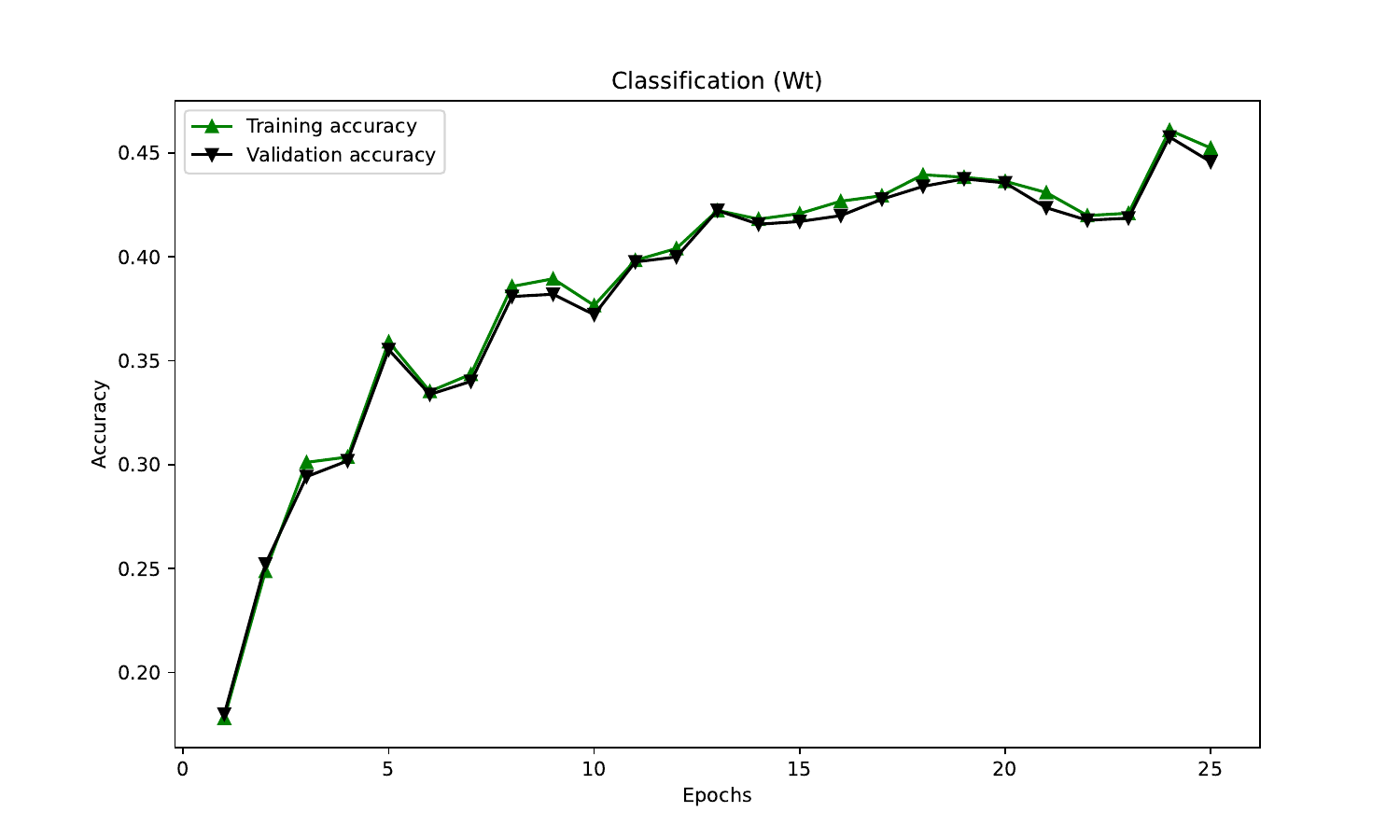}
                    \includegraphics[width=7.5cm,angle=0]{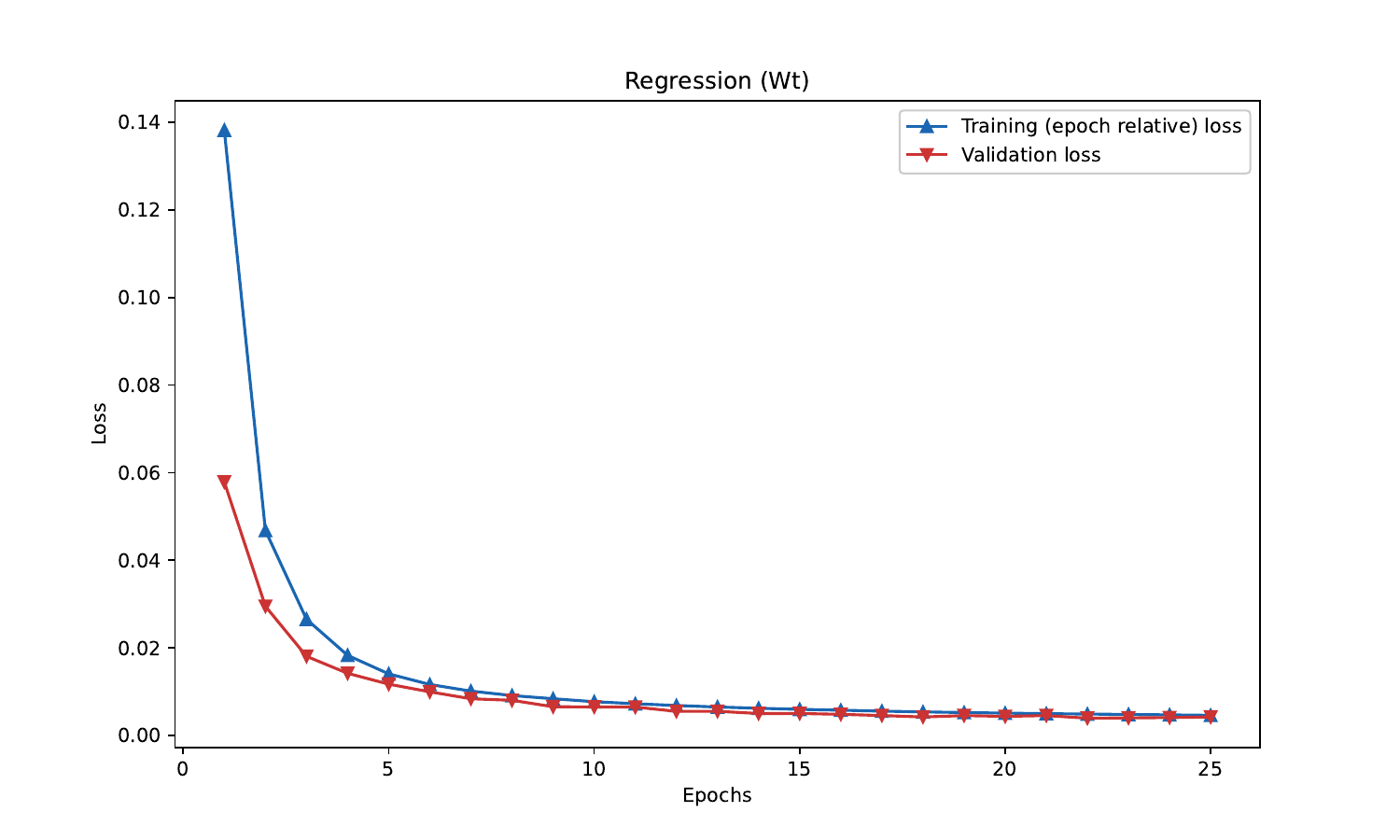}
                    \includegraphics[width=7.5cm,angle=0]{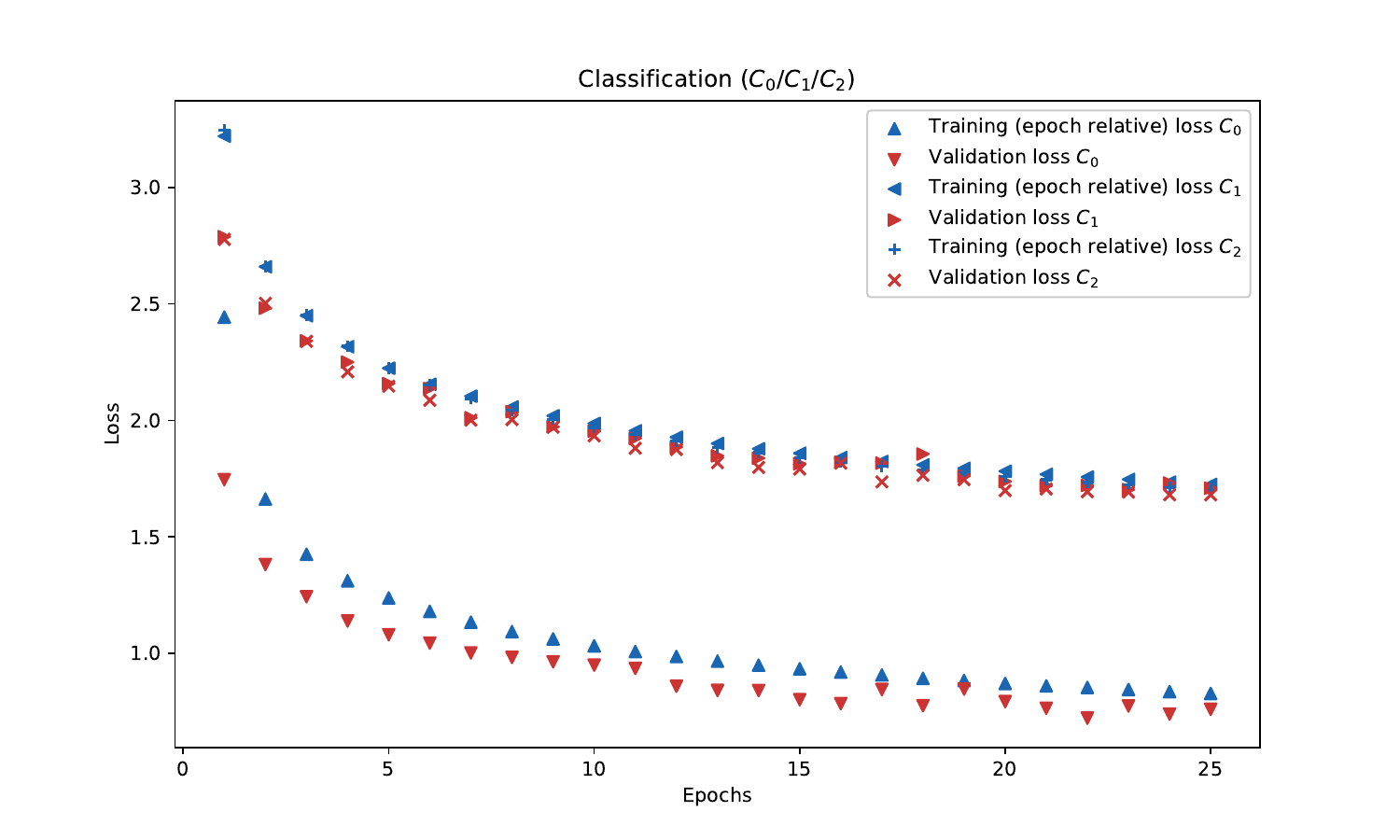}
                    \includegraphics[width=7.5cm,angle=0]{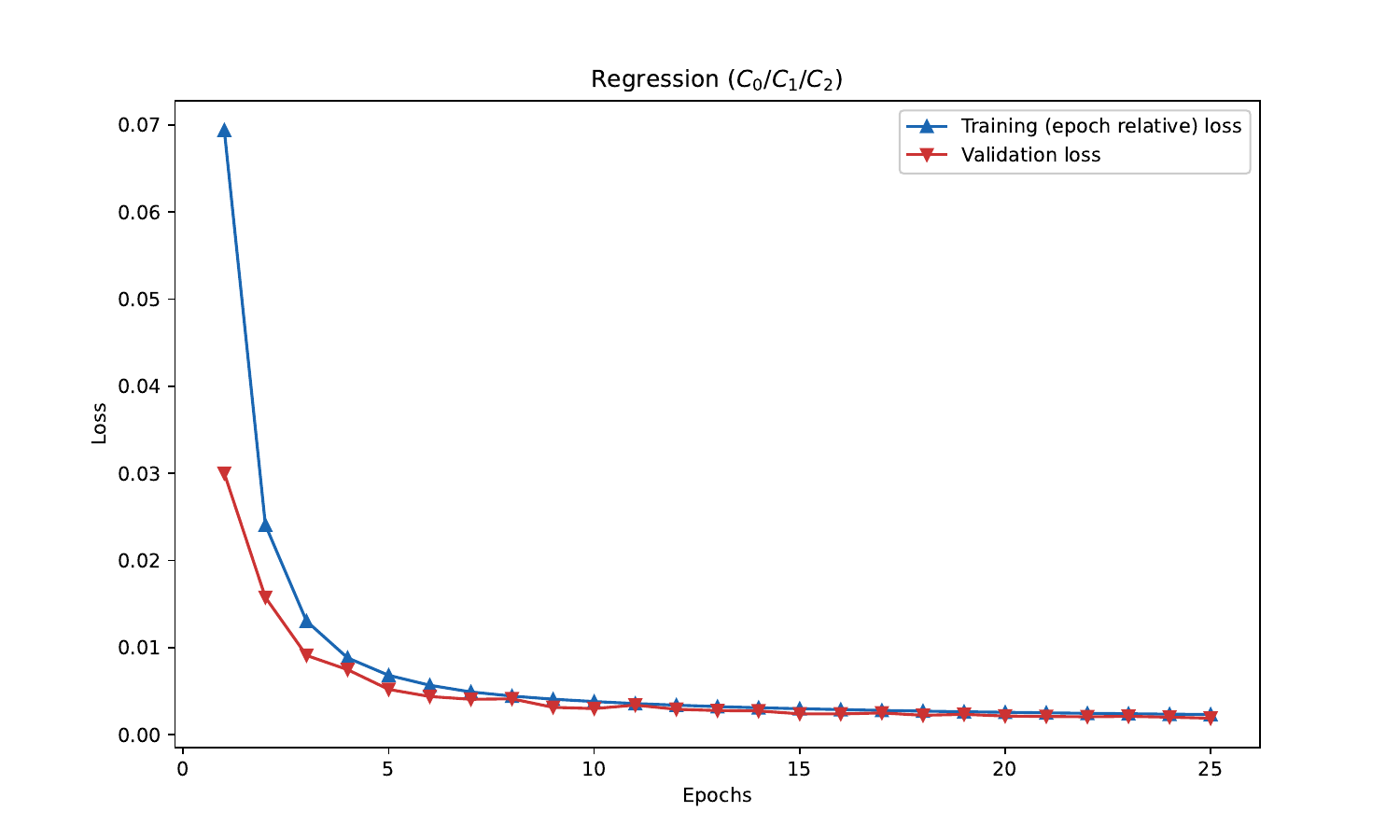}
                    \includegraphics[width=7.5cm,angle=0]{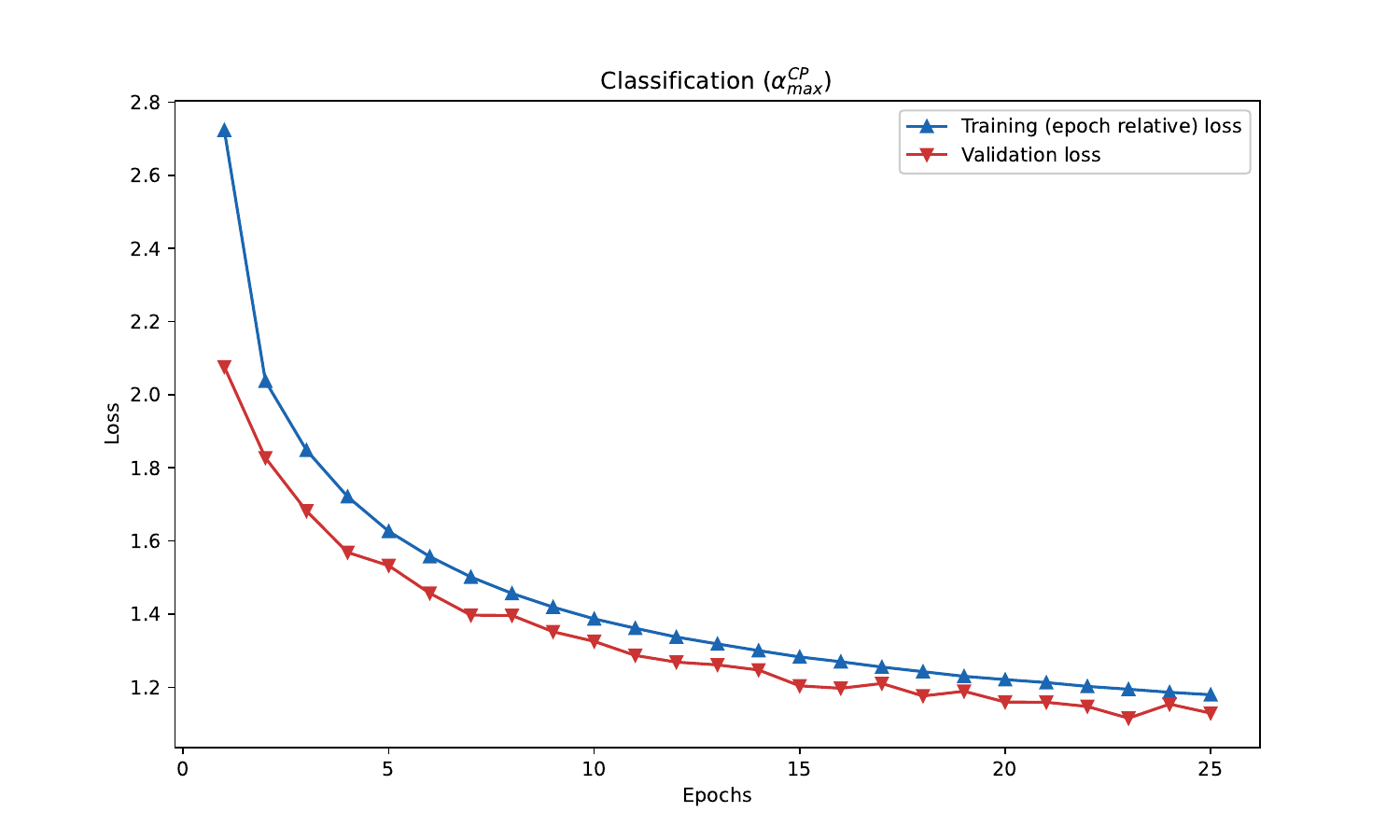}       
                    \includegraphics[width=7.5cm,angle=0]{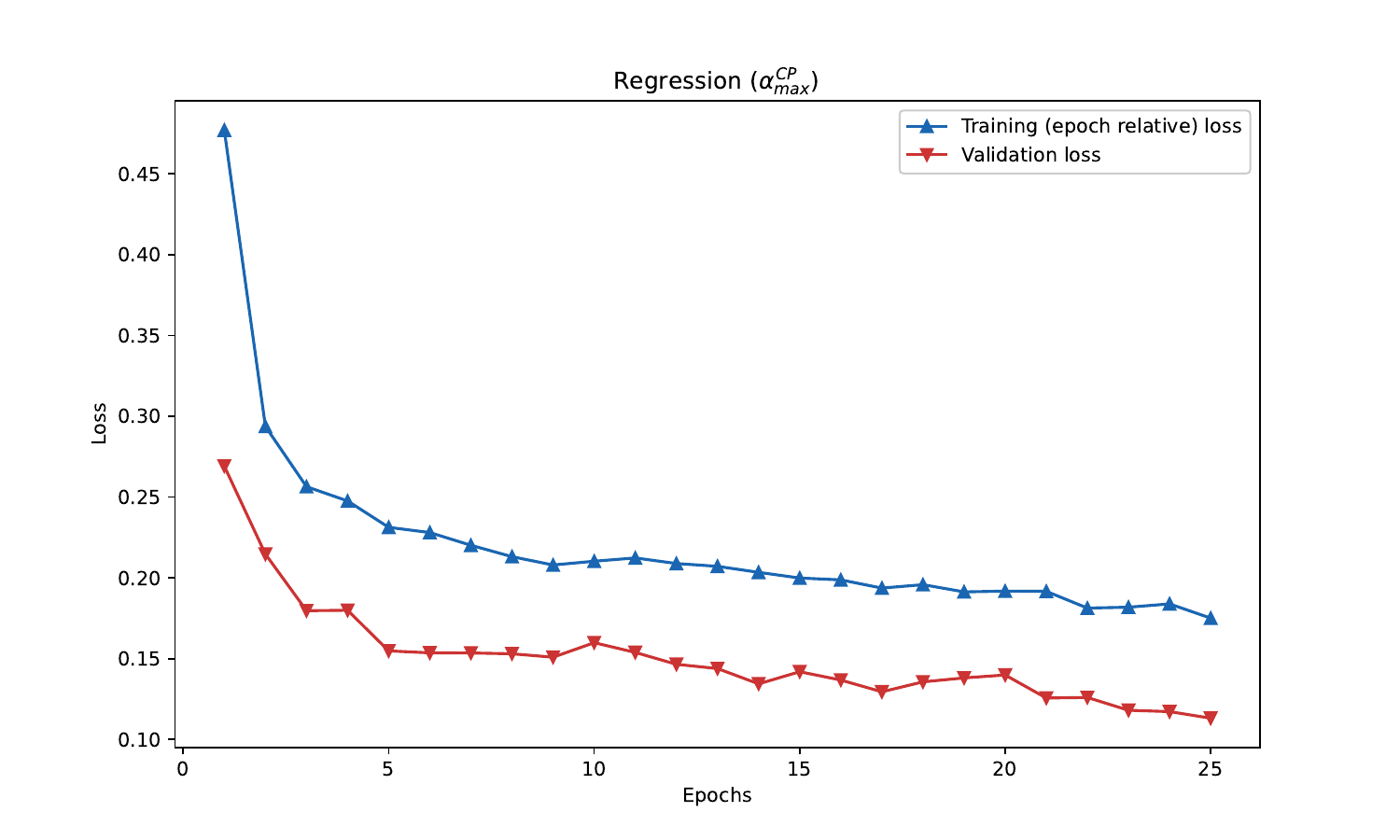}
                    }
                    \end{center}
                    \caption{\raggedright The DNN performance observed during the training process for 
                    {\it Variant-All}: accuracy for {\tt Classification:weights} and loss for all 
                    the other models. For the classification, $N_{class}=51$ classes were used. 
                    \label{figApp:DNN_training}}
                \end{figure}

                Convergence can be observed for all the configurations trained. Figure 
                \ref{figApp:DNN_training} shows it for models trained on the {\it Variant-All} 
                feature set.

            \subsection{Unweighted events} 
                Event weights were stored as a two-dimensional matrix, in
                which rows represented particular events and columns represented 
                $\alpha^{CP}_{max}$ hypotheses. We transformed the matrix into the one 
                containing ones and zeroes by using a Monte Carlo approach: if an element was 
                greater than or equal to a number generated using a uniform random generator, 
                it was replaced by $1$ or $0$, otherwise.

                Therefore, each column of the unweighted events matrix contained
                those events that statistically represented the corresponding hypothesis. The 
                columns were then used as a mask for filtering all the events and getting only 
                the ones belonging to a specified $\alpha^{CP}_{max}$ hypothesis class. The fact 
                that the summed distribution of the filtered events depicted the chosen 
                hypothesis can be observed in Figures \ref{fig:DNN_test_All} - 
                \ref{fig:DNN_test_1.1}.

            \subsection{Negative weights}
                DNN models made predictions on the unweighted events, but due to the lack of 
                restrictions in some of the configurations regarding the sign of the obtained 
                $\alpha^{CP}_{max}$ weights, some of the predictions (up to 10\%) contained 
                negative weights. Such predictions did not have any physical interpretation, 
                as the models were supposed to provide probabilities. We rejected those events 
                containing negative values, ensuring they were added to the summed distribution 
                with zero weight. Vectors containing the summed distribution were then normalised 
                to allow us to compare predictions with true values on the same diagram.
    \end{appendices}

    \clearpage
    \printbibliography
\end{document}